\documentclass[journal,onecolumn]{IEEEtran}

\usepackage{cite}
\usepackage{amsmath,amssymb,amsfonts}
\usepackage{algorithmic}
\usepackage{graphicx}
\usepackage{textcomp}
\usepackage{cite}
\usepackage{mciteplus}
\usepackage{graphicx}
\usepackage{bm}
\usepackage{xcolor}
\usepackage{acro} 

\DeclareAcronym{RIS}{
  short = RIS ,
  long  = reconfigurable intelligent surfaces ,
  class = abbrev
}
\DeclareAcronym{6G}{
  short = 6G,
  long  = sixth generation ,
  class = abbrev
}
\DeclareAcronym{5G}{
  short = 5G,
  long  = fifth generation ,
  class = abbrev
}
\DeclareAcronym{EM}{
  short = EM ,
  long  = electromagnetic,
  class = abbrev
}
\DeclareAcronym{BS}{
  short = BS,
  long  = base station ,
  class = abbrev
}
\DeclareAcronym{UE}{
  short = UE ,
  long  = user equipment ,
  class = abbrev
}
\DeclareAcronym{MISO}{
  short = MISO,
  long  = multiple-input single-output ,
  class = abbrev
}
\DeclareAcronym{MMSE}{
  short = MMSE ,
  long  = minimum mean squared error ,
  class = abbrev
}
\DeclareAcronym{DFT}{
  short = DFT,
  long  = discrete Fourier transform ,
  class = abbrev
}
\DeclareAcronym{THz}{
  short = THz,
  long  = Terahertz ,
  class = abbrev
}
\DeclareAcronym{IoT}{
  short = IoT,
  long  = internet of things ,
  class = abbrev
}
\DeclareAcronym{MSE}{
  short = MSE,
  long  = mean square error ,
  class = abbrev
}
\DeclareAcronym{CSI}{
  short = CSI ,
  long  = channel state information ,
  class = abbrev
}
\DeclareAcronym{MIMO}{
  short = MIMO,
  long  = multiple-input multiple-output ,
  class = abbrev
}
\DeclareAcronym{UPA}{
  short = UPA,
  long  = uniform planner array ,
  class = abbrev
}
\DeclareAcronym{RF}{
  short = RF,
  long  = radio-frequency ,
  class = abbrev
}
\DeclareAcronym{mmWave}{
  short = mmWave,
  long  = millimeter-wave ,
  class = abbrev
}
\DeclareAcronym{AoA}{
  short = AoA ,
  long  = angle of arrival ,
  class = abbrev
}
\DeclareAcronym{AoD}{
  short = AoD,
  long  = angle of departure ,
  class = abbrev
}
\DeclareAcronym{EKF}{
  short = EKF,
  long  = extended Kalman filter ,
  class = abbrev
}
\DeclareAcronym{LMS}{
  short = LMS,
  long  = least mean square ,
  class = abbrev
}
\DeclareAcronym{BiLMS}{
  short = BiLMS,
  long  = bi-directional LMS ,
  class = abbrev
}
\DeclareAcronym{SNR}{
  short = SNR,
  long  = signal-to-noise ratio ,
  class = abbrev
}
\DeclareAcronym{LoS}{
  short = LoS,
  long  = line-of-sight ,
  class = abbrev
}
\DeclareAcronym{TDD}{
  short = TDD,
  long  = time-division duplexing ,
  class = abbrev
}
\DeclareAcronym{NMSE}{
  short = NMSE,
  long  = normalized mean square error ,
  class = abbrev
}
\DeclareAcronym{SDR}{
  short = SDR,
  long  = semidefinite relaxation ,
  class = abbrev
}
\DeclareAcronym{QoS}{
  short = QoS,
  long  = quality of service ,
  class = abbrev
}
\DeclareAcronym{NOMA}{
  short = NOMA,
  long  = non-orthogonal multiple access ,
  class = abbrev
}
\DeclareAcronym{OFDM}{
  short = OFDM,
  long  = orthogonal frequency division multiplexing ,
  class = abbrev
}
\DeclareAcronym{OMA}{
  short = OMA,
  long  = orthogonal multiple access ,
  class = abbrev
}
\DeclareAcronym{NU}{
  short = NU,
  long  = near user ,
  class = abbrev
}
\DeclareAcronym{FU}{
  short = FU,
  long  = far user ,
  class = abbrev
}
\DeclareAcronym{SIC}{
  short = SIC,
  long  = successive interference cancellation ,
  class = abbrev
}
\DeclareAcronym{PLS}{
  short = PLS,
  long  = physical layer security ,
  class = abbrev
}
\DeclareAcronym{MRT}{
  short = MRT,
  long  = maximum ratio transmission ,
  class = abbrev
}
\DeclareAcronym{AWGN}{
  short = AWGN,
  long  = additive white Gaussian noise,
  class = abbrev
}
\DeclareAcronym{SINR}{
  short = SINR,
  long  = signal-to-interference-plus-noise ratio ,
  class = abbrev
}
\DeclareAcronym{BPSK}{
  short = BPSK,
  long  = binary phase shift keying ,
  class = abbrev
}
\DeclareAcronym{QPSK}{
  short = QPSK,
  long  = quadrature phase shift keying ,
  class = abbrev
}
\DeclareAcronym{SVD}{
  short = SVD,
  long  = singular value decomposition ,
  class = abbrev
}
\DeclareAcronym{EVD}{
  short = EVD,
  long  = eigenvalue decomposition ,
  class = abbrev
}
\DeclareAcronym{PDF}{
  short = PDF,
  long  = probability density function ,
  class = abbrev
}
\DeclareAcronym{SER}{
  short = SER,
  long  = symbol error rate ,
  class = abbrev
}
\DeclareAcronym{MGF}{
  short = MGF,
  long  = moment generating function ,
  class = abbrev
}
\DeclareAcronym{2D}{
  short = 2D,
  long  = two-dimensional ,
  class = abbrev
}
\DeclareAcronym{3D}{
  short = 3D,
  long  = three-dimensional ,
  class = abbrev
}
\DeclareAcronym{CLT}{
  short = CLT,
  long  = central limit theorem ,
  class = abbrev
}
\DeclareAcronym{QAM}{
  short = QAM,
  long  = quadrature amplitude modulation ,
  class = abbrev
}
\DeclareAcronym{SISO}{
  short = SISO,
  long  = single-input single-output ,
  class = abbrev
}
\DeclareAcronym{CE}{
  short = CE,
  long  = channel estimation ,
  class = abbrev
}
\DeclareAcronym{KG}{
  short = $K_G$,
  long  = generalized-K ,
  class = abbrev
}
\DeclareAcronym{LSKRF}{
  short = LSKRF,
  long  = least squares Khatri-Rao factorization ,
  class = abbrev
}
\DeclareAcronym{IFFT}{
  short = IFFT,
  long  = inverse fast Fourier transform ,
  class = abbrev
}
\DeclareAcronym{FFT}{
  short = FFT,
  long  = fast Fourier transform ,
  class = abbrev
}
\DeclareAcronym{CP}{
  short = CP,
  long  = cyclic prefix ,
  class = abbrev
}
\DeclareAcronym{ISI}{
  short = ISI,
  long  = inter-symbol interference ,
  class = abbrev
}
\DeclareAcronym{CIR}{
  short = CIR,
  long  = channel impulse response ,
  class = abbrev
}
\DeclareAcronym{LS}{
  short = least square,
  long  = channel impulse response ,
  class = abbrev
}
\DeclareAcronym{CFR}{
  short = CFR,
  long  = channel frequency response ,
  class = abbrev
}
\DeclareAcronym{OTFS}{
  short = OTFS,
  long  = orthogonal time frequency space ,
  class = abbrev
}
\DeclareAcronym{BER}{
  short = BER,
  long  = bit error rate ,
  class = abbrev
} 
\DeclareAcronym{PAPR}{
  short = PAPR,
  long  = peak-to-average power ratio ,
  class = abbrev
}
\DeclareAcronym{GFDM}{
  short = GFDM,
  long  = generalized frequency division multiplexing ,
  class = abbrev
}
\DeclareAcronym{CFO}{
  short = CFO,
  long  = carrier frequency offset ,
  class = abbrev
}
\DeclareAcronym{CCDF}{
  short = CCDF,
  long  =complementary cumulative distribution function ,
  class = abbrev
}
\DeclareAcronym{ICI}{
  short = ICI,
  long  = intercarrier-interference ,
  class = abbrev
}

\DeclareAcronym{VSA}{
  short = VSA,
  long  = vector signal analyzer,
  class = abbrev
}

\usepackage{subfigure}

\usepackage{amsmath}
\usepackage{caption}
\usepackage{graphicx}   
\usepackage{tabularx}
\usepackage{breqn}
\usepackage{xspace} 
\usepackage{algorithm,algorithmic}

\usepackage{cellspace}
\usepackage{tabularx,booktabs,caption,ragged2e}
\usepackage{amsmath}
\usepackage{epsfig}
\usepackage{cite}
\usepackage{mciteplus}
\usepackage{graphicx}
\usepackage{bm}
\usepackage{xcolor}
\usepackage{amsmath}

\usepackage{caption}
\usepackage{algorithm,algorithmic}
\usepackage{graphicx}
\usepackage[font=small,labelfont=bf]{caption}
\usepackage{cite}
\usepackage{mciteplus}
\usepackage{breakcites}
\usepackage{amsthm}
\usepackage{graphicx}
\usepackage{epsfig}
\usepackage{psfrag}
\usepackage{epstopdf}
\usepackage{subfigure}
\usepackage{amsmath}

\usepackage{stfloats}
\usepackage{amsmath}
\usepackage{amsfonts}
\usepackage{amssymb}
\usepackage{gensymb}
\usepackage{eqnarray}
\usepackage{subeqnarray}

\usepackage{array}

\usepackage{flushend}
\usepackage{setspace}
\usepackage{paralist}

\usepackage{suffix}
\usepackage{mathtools}
\usepackage{float}

\def\BibTeX{{\rm B\kern-.05em{\sc i\kern-.025em b}\kern-.08em
    T\kern-.1667em\lower.7ex\hbox{E}\kern-.125emX}}
\begin{document}
\title{Experimental emulation for OTFS waveform RF-impairments }

\author{Abdelrahman Abushattal, {\it Student Member, IEEE}\thanks{A. Abushattal and Ayhan Yazgan are with the Department of Electrical and Electronics Engineering, Karadeniz Teknik Üniversitesi, Trabzon, Turkey (e-mail:ceabushattal@gmail.com  \& : ayhanyazgan@ktu.edu.tr arslan@usf.edu.)}, Salah Eddine Zegrar, {\it Student Member, IEEE}\thanks{S. Zegrar and H. Arslan are with the Department of Electrical and Electronics Engineering, Istanbul Medipol University, Istanbul, Turkey (e-mail:salah.zegrar@std.medipol.edu.tr  \& arslan@usf.edu.)}, Ayhan Yazgan, {\it Senior Member, IEEE}
and H\"{u}seyin Arslan, {\it Fellow Member, IEEE}}

\maketitle

\begin{abstract}
Orthogonal time-frequency space (OTFS) waveform exceeds the challenges that face orthogonal frequency division multiplexing (OFDM) in the high-mobility environment with high time-frequency dispersive channels. Since practical pulse shaping design and RF-impairments effects have a direct impact on waveform behavior, this paper investigates experimental implementation for practical pulse shaping design and RF-impairments that affect OTFS waveform performance and compares them to OFDM waveform as a benchmark. Firstly, the doubly-dispersive channel effect is analyzed, then an experimental framework is established for investigating the  RF-impairments include non-linearity, Carrier frequency offset,  I/Q-imbalances, DC-offset, and phase noise  are considered. The experiments were conducted in a real indoor wireless environment using software-defined radio (SDR)based on the Keysight EXG X-Serie devices. The experimental results validate the accuracy of the theoretical results.
\end{abstract}



\maketitle

\section{Introduction}
\label{sec:introduction}
\begin{table}[!h]
 \caption {NOMENCLATURES}
\renewcommand{\arraystretch}{0.5}
\resizebox{0.6\columnwidth}{!}{
\begin{tabular}{l l }
\hline
{\textbf{Abbreviation }}                                                                         & {\textbf{Definition}} \\ \hline
{5G}                                                  & Fifth Generation                           \\ 
{6G}                                                  & Sixth Generation                           \\ 
 {A/D}   & Analogto-Digital                \\
{BER}                                              & Bit Error Rate                           \\ 
  {CCDF}  & Cumulative Distribution Function             \\
   {CFO}    & Carrier Frequency Offset                 \\ 
   {CP}    & Cyclic Prefix                      \\
 {D/A}    &Digital-To-Analog()              \\
    {DC}    & Direct Current             \\
{eMBB}             &  Enhanced-Mobile Broadband     \\ 
  {I/Q imbalance}    &  In-Phase And Quadrature Imbalance            \\
  {ICF}    & Iterative Clipping And Filtering                       \\ 
{ICI  } & Iter-Carrier Interference      \\
     {ISFFT}    &     inverse symplectic finite Fourier transform                 \\
{ISI}    & Inter-Symbol Interference  \\ 
{JRC }                                             & Joint Radar Communication     \\ 
 {LNA} & Low-Noise Amplifier                \\
     {LS}    & Least Squares                  \\
    {MIMO}    & Multi-Input Multi Output               \\
{mMTC}           &  Massive Machine Type Communications               \\ 
{mmWave}    & millimeter wave                   \\
    {MP}    & Message Passing                   \\
{NOMA }    & Non-Orthogonal Multiple Access                          \\ 
{OFDM}          & Orthogonal Frequency Division Multiplexing   \\
{OTFS}                      & Orthogonal Time-Frequency Space \\
{PA}    & Power Amplifier                \\
  {PAPR}  & Peak-To-Average Power Ratio   \\ 
 {PSW}    & Prolate Spheroida Waveform    \\            
     {QAM}    & Quadrature Amplitude Modulation                   \\
{RF}    & Radio-Frequency                 \\
{SDR}                 & Software-Defined Radio    \\ 
     {SFFT}    &   Symplectic Finite Fourier Transform                 \\
    {URLLC}    & Ultra-Reliable And Low Latency Communications        \\ 
    {VSA}    & Vector Signal Analyzer                    \\
   {VSG}    & Vector Signal Generator                    \\
{RCF}    & Raised-cosine filter                    \\
{LO}    & Local oscillator                     \\
{HPA}    & high power amplifier                   \\
{OOB}    & out-of-band                \\
{AWGN}    & additive white Gaussian noise               \\
{CMOS}    & complementary metal-oxide-semiconductor     \\
{VCO}    & voltage control oscillator     \\

    \\ 
  \hline
\end{tabular}}

\end{table}
\IEEEPARstart {T}{he} exponential growth in the number of connected devices created an urge to differentiate and fulfill the various requirements of the users in the network so that all are served properly\cite{index2019forecast}. These needs are the main driving factors of the \ac{5G}, that mainly support three services i.e., enhanced-mobile broadband (eMBB), ultra-reliable and low latency communications (URLLC), and massive machine type communications (mMTC). These different services are achieved by using multiple \ac{OFDM} numerologies \cite{Arslan-5G,jaradat1}. The same concept is visioned for the \ac{6G}, where instead of implementing \ac{OFDM} with multiple parameters, the network will be ultra-flexible and will accommodate multiple different waveforms in a single frame to meet the requirements of the \ac{6G} \cite{Arslan-6G}.  

One of the most promising waveform candidates is the \ac{OTFS} waveform, that represents the information symbols in the delay-Doppler domain, where all modulated data experiences almost the same channel gain even at high mobility cases \cite{3}. This enhances the performance of the system that suffers from high Doppler frequency shift compared to conventional multi-carrier techniques such as \ac{OFDM} \cite{4}.

The rich scattering environment, and the mobility of transmitter, receiver, or scatterers lead to fast variation in time/frequency response of the wireless channel, which is very hard and expensive to estimate and compensate \cite{2}. 
while, the \ac{OTFS} combats all of these channel effects better than most conventional schemes in high time/frequency dispersion channel, beside, it has been shown that \ac{OTFS} achieves better \ac{BER} performance compared to \ac{OFDM} for mobile user with velocity ranging between 30 and 500 Km/h \cite{3,4}.\par
According to above features, \ac{OTFS} gained interest in various applications. Lately, in \cite{8835764,gaudio2020effectiveness}, \ac{OTFS} was investigated for joint radar communication (JRC) system. In\cite{8835764}, authors proposed \ac{OTFS}-based matched filter where it is shown that \ac{OTFS} provide better tracking speed and radar distance range compare to \ac{OFDM} waveform. while in \cite{gaudio2020effectiveness}, authors examined OTFS performances in vehicle applications for mono-static radar. In \cite{8786203,9354639} non-orthogonal multiple access (NOMA) integrated with \ac{OTFS} to provide spectral efficiency and serve  multiple users with different mobility characteristic (i.e. stationary and high mobility) in heterogeneous networks. In \cite{8058662,8580850,8746382}, author investigate  the performance of \ac{OTFS} in mm-Wave communications. it is shown that \ac{OTFS} provides better robustness against high Doppler shift and phase noise that exists in mmWave communications. However, the effect of the waveform physical design should be considered to have a clear understanding of the system's performance. Therefore, the effect of the practical pulse shaping design and RF-impairments should be described according to experimental evaluation. the following section explains the related research that deals with pulse shaping design and RF-impairments.

\subsection{Related literature }
Even though the \ac{OTFS}  works well in demanding channel circumstances, \ac{RF} impairments should be investigated, validated, and compared to traditional waveform schemes. 

Regarding the \ac{PAPR} of the OTFS had been analyzed in \cite{PAPR} and its performance was compared to \ac{OFDM} \cite{jaradat2} waveform. The results showed that \ac{OTFS} has a better \ac{PAPR} performance that has a linear relationship with the number of Doppler bins. To minimize the PAPR in a pilot embedded OTFS, a modified iterative clipping, and filtering(ICF) was proposed in \cite{gao2020peak}.\\
The I/Q imbalance impairments were discussed in \cite{IQ} for \ac{OTFS} systems, where it was shown that the \ac{BER} curve is saturating beyond certain \ac{SNR}, which clearly indicates that I/Q imbalance needs to be compensated.
In order to implement the \ac{OTFS} and study the receiver impairments effects, software-defined radio (SDR) is used. SDR is a radio communication system where all or most of the physical layer functions have been implemented in software. As discussed in \cite{5}, the authors implement an SDR design for the \ac{OTFS} modem, and investigate the \ac{CFO} and direct current (DC) offset impairments for the real indoor wireless channel. However, in this work, no mobility was considered in the experiment.

\subsection{Motivation and Contribution}
The waveform's behavior is directly impacted by practical pulse shaping design and RF-impairments effects. 
Furthermore, radio frequency (RF) impairments including as non-linearity, Doppler dispersion, I/Q-imbalances, DC-offset, and phase noise have been shown to significantly degrade the performance of the waveform.
Since \ac{OTFS} is the promising waveform for the communication under high mobility conditions and exhibits resilience to narrow-band interference, the degradation due to the design on the \ac{OTFS} system should be known to provide the optimal system design. Also, taking into consideration one \ac{RF}-impairment when designing an \ac{OTFS} communication system may increase the effect of another impairment. Therefore, the \ac{OTFS} system should be investigated under different \ac{RF}-impairments.
Thus, Understanding the RF-impairments effects on the system performance is one of the significant and critical. Besides, to get a clear view of the OTFS waveform’s performance, it is necessary to carry out an experimental study of the system.

\begin{figure*}
  \centering
\includegraphics[width=\textwidth , height=4.55cm]{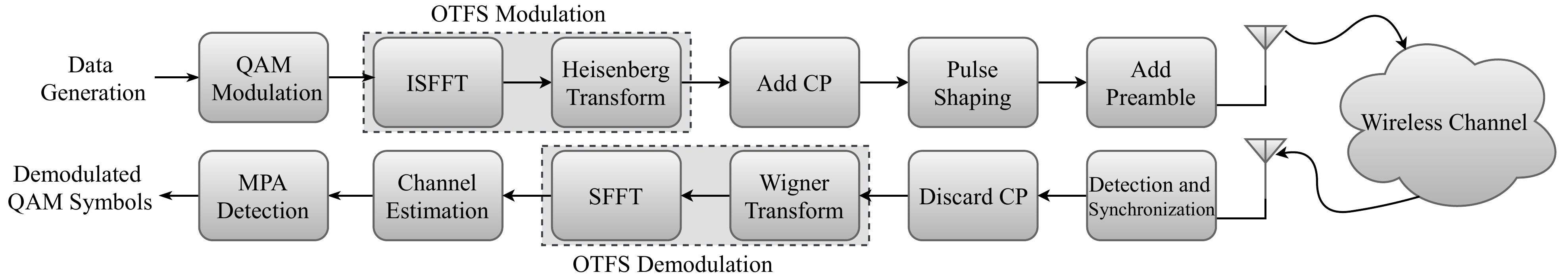}
  \caption{OTFS transceiver block diagram.}
  \label{fig:transceiver}
\end{figure*}
 The main contributions of this paper are summarized is to present and emulates different \ac{RF}-impairments at both transmitter and receiver when \ac{OTFS} waveform is used for communication, and then, it compares these impairments to \ac{OFDM} waveform. This work concentrate on major RF front-end impairments: \ac{PAPR}, \ac{CFO}, I/Q imbalance, DC-offset, and phase noise impairments. Testing and evaluation of the \ac{RF}-impairments of the \ac{OTFS} and \ac{OFDM} waveforms are examined under real-time experiment using Keysight Agilent Technologies EXA signal analyzer N9010A \cite{EXA}.

\subsection{Organization}
The remainder of this paper is organized as follows. Section \ref{Sec:modulation} presents the modulation and demodulation of both {OTFS} and {OFDM} waveforms. 

Section \ref{Sec:Impairement} discusses the channel effects and the RF-impairments impact, and shows the real implementation results. Finally, Section \ref{Sec:Conclusion} concludes the work.

\section{Waveform Modulation and Demodulation}\label{Sec:modulation}
The waveform is known as the physical shape of information represented by a signal transmitted through the channel. The transmitted signal $x(t)$ in a pulse-shaping system is formed by modulating data symbols $d_{n, k}$ onto time-frequency (Delay-Doppler) shifted versions of a transmit pulse $g(t)$ i.e.,
\begin{equation}
x(t)=\sum_{n \in \mathbb{Z}} \sum_{k \in \mathbb{Z}} d_{n, k} g_{n, k}(t),
\label{equ:Tx-pulse}
\end{equation}
with
\begin{equation} \label{filterTx}
g_{n, k}(t)=\left(\mathbb{M}_{k F} \mathbb{D}_{n T} g\right)(t)=g(t-n T) e^{j 2 \pi k F t},
\end{equation}
where $\mathbb{M}_{\nu} \mathbb{D}_{\tau}$ is the time-frequency shift operator that includes a delay (time shift) $\tau=n T$ and a modulation (frequency shift) $\nu=k F$, $n$ is the time index, $k$ is the subcarrier index, $T$ is the sampling period, and $F$ is the sub-carrier spacing.

\subsection{OFDM Waveform}

In {OFDM} systems, $N$ data symbols $X(k), k=0, 1, ..., N-1$ are mapped in the frequency domain i.e., $\mathbb{D}_{\tau} =0$. The pulse shaping filter $g(t)$ has a rectangular pulse shape in the transmitted lattice \cite{sahin2013survey}. Then, by using \eqref{equ:Tx-pulse}, the transmitted {OFDM} discrete time-domain signal is given by
\begin{equation}
x[n]=\frac{1}{\sqrt{N}} \sum_{k=0}^{N-1} X(k) e^{j 2 \pi n k / N}.
\label{equ:Tx-OFDM}
\end{equation}

The discrete signal expression of the $n$-th received sample is given as
\begin{equation}
y[n]=\frac{1}{\sqrt{N}} \sum_{k=0}^{N-1} H(k)X(k) e^{j 2 \pi n(k) / N}+w[n],
\label{equ:Rx-OFDM}
\end{equation}
where ${ H(k)}$ denotes the channel coefficients, and $w(n)$ is the zero-mean {AWGN} with $\sigma^2$ variance.

\subsection{OTFS Waveform}
{OTFS} modulation is comprised of cascaded two-dimensional (2D) transforms at both transmitter and receiver, as shown in Fig. \ref{fig:transceiver}. At the transmitter side, the information symbols $X[l,k],k=0,..., N-1,l=0,..., M-1$  are map in the two-dimensional delay-Doppler domain from the modulation alphabet $\mathbb{A}$ to be transmit over $N.T$ time duration and using bandwidth $B=M.\Delta f$ . where $\Delta f=1/T$, and $N$,  $M$ are the delay and Doppler bins, respectively. 

Then inverse simplistic finite Fourier transform (ISFFT) using to map the  $N\times M$ delay -Doppler grid points into the time-frequency (TF) plane as follows

\begin{equation}
x[n, m]=\frac{1}{N M} \sum_{k=0}^{N-1} \sum_{l=0}^{M-1} X[k, l] e^{j 2 \pi\left(\frac{n k}{N}-\frac{m l}{M}\right)}.
\label{equ:Tx-OTFS}
\end{equation}

As illustrate by the dashed box in Fig. \ref{fig:transceiver}, the TF plane signal created is transformed to the time domain signal for transmission using the Heisenberg transform, which is given by \cite{3,4}

\begin{equation}
x(t)=\sum_{n=0}^{N-1} \sum_{m=0}^{M-1} X[n, m] g_{t x}(t-n T) e^{j 2 \pi m \Delta f(t-n T)}
\end{equation} 

 where $g_{t x}$ denotes transmit pulse shaping.
 Unlike OFDM where CP  is added for each of $N$ symbols in the frame, the CP is added for each frame in the time domain in OTFS. This considerably reduces the CP overhead. Then the signal will be transmitted through the time-varying wireless channel. The time domain received signal is expressed as

\begin{equation}
y(t)=\int_{\nu} \int_{\tau} h(\tau, \nu) x(t-\tau) e^{j 2 \pi \nu(t-\tau)} \mathrm{d} \tau \mathrm{d} \nu
\end{equation}
where $\tau$ and $\nu$  denote the delay and Doppler variables, respectively. And $h(\tau, \nu)$ represent the complex channel response in delay Doppler domain. 
Wigner transform at the receiver side used to transform the time domain received signal $y(t)$ to TF domain, by matching it with the receiver pulse shaping $g_{rx}$. Assum that the   transmit pulse shaping $g_{t x}$ and the receiver pulse shaping $g_{rx}$ satisfy
bi-orthogonality conditions (The next section will addresses the situation in which the conditions are not satisfied) then the TF signal is given by \cite{3,4,surabhi2019low}

\begin{equation}
Y[n, m]=H[n, m] X[n, m]+W[n, m]
\end{equation}where $W[n,m]$ is the additive white Gaussian noise (AWGN) and $H[n, m]$  is given by 

\begin{equation}
H[n, m]=\int_{\tau} \int_{\nu} h(\tau, \nu) e^{j 2 \pi \nu n T} e^{-j 2 \pi(\nu+m \Delta f) \tau} \mathrm{d} \nu \mathrm{d} \tau
\end{equation}
Then the symplectic finite Fourier transform (SFFT) is used to mapped the TF signal in delay-Doppler domain, which is defined as follows:

\begin{equation}
\begin{aligned}
&\hat y[k, l]=\frac{1}{N M} \sum_{n=0}^{N-1} \sum_{m=0}^{N} Y[n, m] e^{-j 2 \pi\left(\frac{n k}{N}-\frac{m l}{M}\right)} \\
&=\frac{1}{N M} \sum_{n=0}^{N-1} \sum_{m=0}^{M-1} x[n, m]  h(\tau^', \nu^')+w[k, l],
\end{aligned}
\end{equation}


Message passing (MP) detection will be used after OTFS demodulation to detect the received symbols $\hat x[k, l]$ from the Delay Doppler mapped received signal $\hat y[k, l]$, as will be illustrated in the following subsection. This will be done so that the symbols can be extracted from the received signal.

\subsection{Channel Estimation and MP Detection}
\begin{figure*}[!t]
    \begin{center}
    \subfigure[Transmitted symbols.]{\label{TX}\includegraphics[width=60mm]{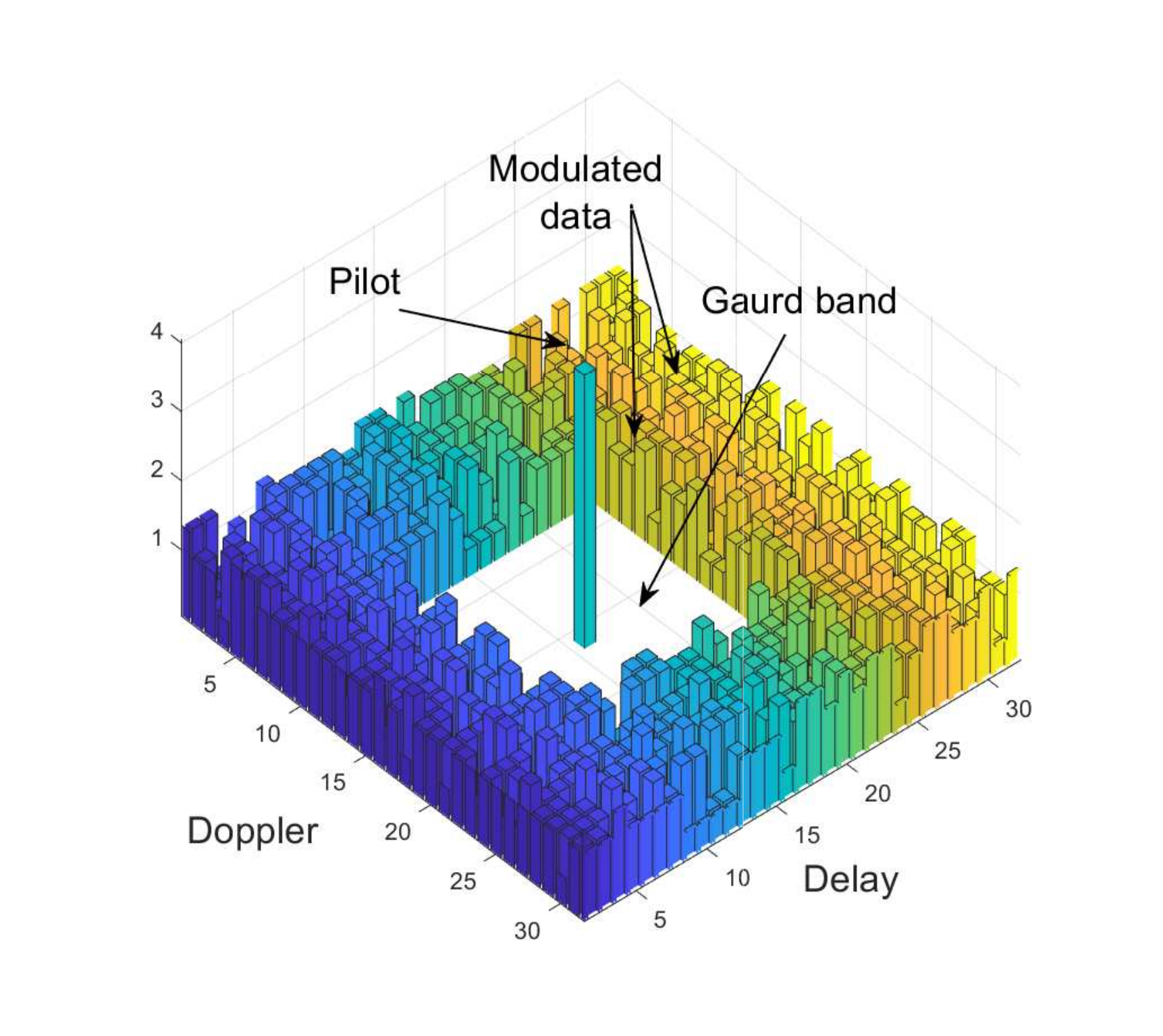}}
    \subfigure[Received signal.]{\label{Rx}\includegraphics[width=60mm]{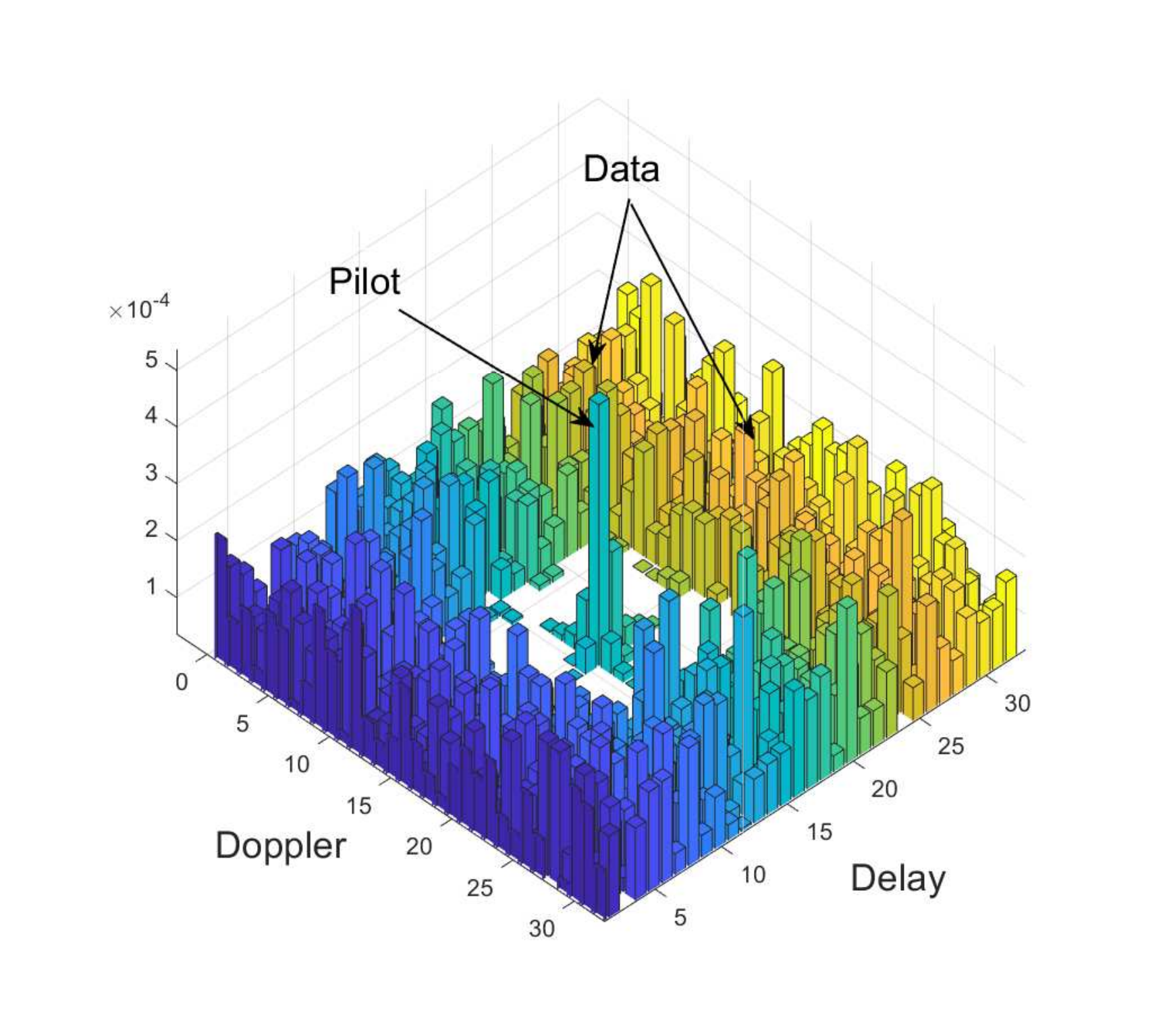}}
    \subfigure[Received embedded pilot.]{\label{RXp}\includegraphics[width=60mm]{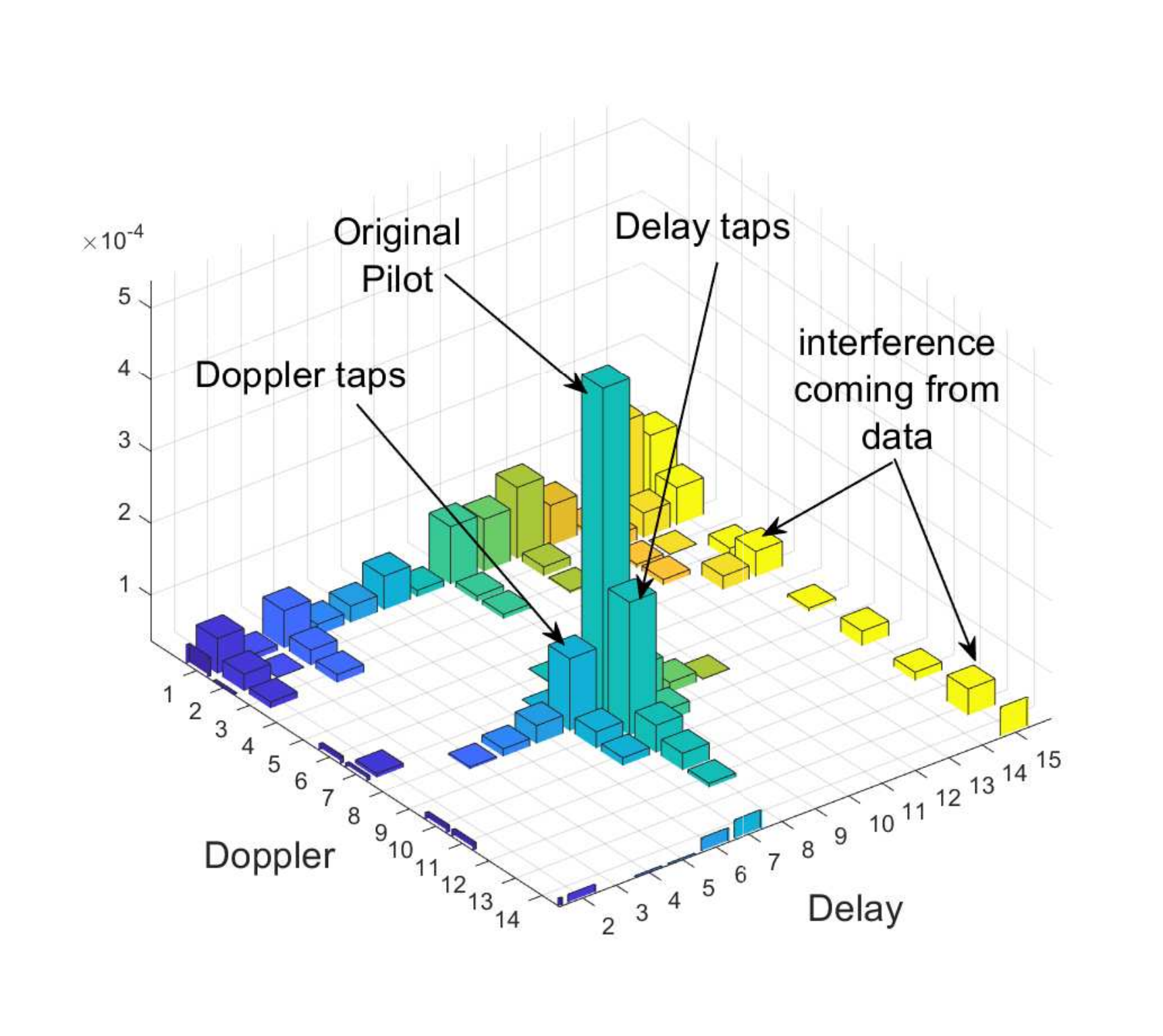}}
    \\
    \end{center}
    \centering
    \caption{The transmitted and received signal structure in the delay-Doppler plane.}
    \label{fig:OTFS_signal}
\end{figure*}
At the receiver side, the doubly-dispersive channel response has to be estimated for the OTFS detection. In \cite{fish2013delay}, the authors proposed a flag method to estimate the channel parameters and improved matched filter algorithm. However, the time-frequency domain estimate of the OTFS channels led to an increase in implementation complexity. In \cite{ramachandran2018mimo}, MIMO-OTFS iterative algorithm was proposed for channel estimation in the delay-Doppler domain, where a whole OTFS frame containing pilots is used to estimate the channel. This estimation is used for next frame data detection. Unfortunately, this approach is valid only for time-invariant channels. In\cite{raviteja2019embedded}, embedded pilot-aided channel estimation scheme is proposed, where the receiver simultaneously proceeds a threshold approach channel estimation followed by message passing (MP) data detection in the same OTFS frame. Encouraged by the latter embedded pilot-aided estimation, in this paper, the OTFS frame is arranged as shown in Fig. \ref{fig:OTFS_signal} (a), where the data is distributed in the delay-Doppler domain. A guard band is adopted to prevent interference between the modulated data and the embedded pilot at the receiver detection as illustrated in Fig. \ref{fig:OTFS_signal} (b). Finally, the delay and Doppler taps (given in Fig. \ref{fig:OTFS_signal} (c)) are estimated and passed to the MP algorithm to detected the transmitted symbols.

After estimating the channel parameters, MP can be used to extract $\hat{x}[k, l]$ from the $\hat{y}[k, l]$   \cite{ramachandran2018mimo,narasimhan2014channel,ge2021receiver,raviteja2018low} as it illustrate in algorithm \ref{MPalgorithm}. The detected received symbols are founded by evaluating the joint maximum a posteriori probability (MAP) represent as follows\cite{8424569}:

\begin{equation}
\hat{x}=\underset{\mathbf{x} \in \mathbb{A}^{N M}}{\arg \max } \operatorname{Pr}({x} \mid {y},{H})
\end{equation}

Adopting the same MAP detection in \cite{8424569}. Let represent $y$ as a complex vector with element represents as $y[d]$ where $1\leq d \leq MN$ and $x$ is the information vector $x[c]$ where $1\leq c \leq MN$ and the complex channel will be represent as $H[d,c]$. Considering symbol by symbol detection $x_c$ and all the transmitted symbols $ x
_c \in \mathbb{A}$ have equivalently probability and independent form ${y}[d]$ , then $\hat{x}[c]$ is represent as     
\begin{equation}
\begin{aligned}
\hat{x}[c]&= \underset{a_{j} \in \mathbb{A}}{\arg \max } \frac{1}{|\mathbb{A}|} \operatorname{Pr}\left(\mathrm{y}[d] \mid x[c]=a_{j}, H[d,c]\right)\\ 
& \approx \underset{a_{j} \in \mathbb{A}}{\arg \max } \prod_{d \in \mathcal{J}_{c}} \operatorname{Pr}\left(y[d] \mid x[c]=a_{j}, H[d,c]\right)
\end{aligned}
\end{equation} 

In MAP detection, the observation node complex vector $y_d$,  is related to the to variable node ${x_e},e \in \mathcal{I}_{d}$. Also, variable node $x_c$ are related to variable nodes ${y_e},e \in \mathcal{J}_{c}$. where $\mathcal{I}_{d}$, $\mathcal{J}_{c}$ represent positions of the $d^{th}$ rows and $c^{th}$ columns of the non zero element in $H$.

As it detailed in \cite{8424569}, massage is passing from observation nodes $y[d]$ to $x[c]$

\begin{equation}
y[d]=x[c] H[d, c]+\underbrace{\sum_{e \in \mathcal{I}(d), e \neq c} x[e] H[d, e]+w[d]}_{\zeta_{d, c}^{(i)}}
\end{equation}

where $\zeta_{d, c}^{(i)}$ is represent the self interference that is could be approximate as Gaussian distribution according to central limit theorem (CLT) by evaluate the mean and variance. For more details see \ref{y_detection} \cite{narasimhan2014channel,ge2021receiver} \cite{raviteja2018low}  , where the approximated means and variance expressed as follow  respectively\cite{8424569}: 

\begin{equation}
 \mu^{(i)}_{d,c}=\mathbb{E}\left(\zeta_{d, c}^{(i)}\right)=\sum_{d \in \mathcal{J}_{c}, e\neq c} h_{d,e}\sum_{j=1, j \neq i}^{\left |\mathbb{A} \right|} \mathbb{E}\left(a_{j}\right) 
\end{equation}
\begin{equation}
=\sum_{e \in \mathcal{I}_{d}, e \neq c} \sum_{j=1}^{|\mathbb{A}|} p_{e d}^{(i)}\left(a_{j}\right) a_{j} H_{d, e}   
\end{equation}
\begin{equation}
{\sigma^{(i)}_{d,c}}^2  =\sum_{e \in \mathcal{I}_{d}, e \neq c}\left(\sum_{j=1}^{|\mathbb{A}|} p_{e d}^{(i)}\left(x_{j}\right)\left|x_{j}\right|^{2}\left|H_{d, e}\right|^{2}-\left|\mu_i\right| ^2\right)
\end{equation}where the pmf $p_{c d}^{(i)}$ is evaluated as following 
\begin{equation}
p_{c d}^{(i)}\left(x_{j}\right)=\Delta \cdot \tilde{p}_{c d}^{(i)}\left(x_{j}\right)+(1-\Delta) \cdot p_{c d}^{(i-1)}\left(x_{j}\right)
\end{equation}
where  $\Delta$ is the damping factor that determine the weight of the previous probability distribution to evaluate the next one and it value in range $0< \Delta \leq 1$ \cite{pretti2005message}, and

\begin{equation}
\tilde{p}_{c d}^{(i)}\left(a_{j}\right) \propto \prod_{e \in \mathcal{J}_{c}, e \neq d} \operatorname{Pr}\left(y_{e} \mid x_{c}=a_{j}, \mathbf{H}\right)
\end{equation}where 
\begin{equation}
\operatorname{Pr}\left(y_{e} \mid x_{c}=a_{j}, \mathbf{H}\right) \propto \exp \left(\frac{-\left|y_{e}-\mu_{e c}^{(i)}-H_{e, c} a_{j}\right|^{2}}{\sigma_{e c}^{2,(i)}}\right)
\end{equation}
The decisions made upon that symbols to be transmitted are set as follows:
\begin{equation}
\hat{x}_{c}=\underset{a_{j} \in \mathbb{A}}{\arg \max } p_{c}\left(a_{j}\right), \quad c \in\{0, \cdots, N M-1\}
\end{equation}where
$
p_{c}\left(a_{j}\right)=\prod_{e \in \mathcal{J}_{c}} \operatorname{Pr}\left(y_{e} \mid x_{c}=a_{j}, \mathbf{H}\right)$

\begin{algorithm}[h!]
\begin{algorithmic}[1]

\item Input parameters:\\
 $\textbf{H};\textbf{y}; N; M; M_{mod};$and$ n_itr $; 
    \item Initialization :
$i=0$ and $\left|\mathbf{p}_{c d}^{(0)}=1 /\right| \mathbb{A} \mid$ for $c=\{0, \cdots, N M-1\}$ and $d \in \mathcal{J}_{c}$
$\mu_{i}=0; \sigma^2_{i}=0; $

\item{ Procedure:
 
  \FOR{ i=1:\textbf{Length}(iteration)} 
  
  \FOR{ m=1:M } 
  \FOR{ n=1:N } 
\item \textbf{Compute and update}\\
  $\mu_{mn}^{(i)}; \left(\sigma_{m n}^{(i)}\right)^{2} $ 
    \ENDFOR 
    \ENDFOR

   \FOR{ each $ d \in \mathcal{J}_{c}$ } %
 
 \item \textbf{Compute}\\
 $p_{c d}^{(i+1)}\left(a_{j}\right)=\Delta \cdot p_{c d}^{(i)}\left(a_{j}\right)+(1-\Delta) \cdot p_{c d}^{(i-1)}\left(a_{j}\right)$
 \\
 \item \textbf{Compute} $p_{c d}^{(i)}\left(a_{j}\right) \propto \prod_{e \in \mathcal{J}_{c}, e \neq d} \operatorname{Pr}\left(y_{e} \mid x_{c}=a_{j}, \mathbf{H}\right)$,
\ENDFOR 
  
  \ENDFOR
  \IF{$\max _{c, d, a_{j}}\left|p_{c d}^{(i+1)}\left(a_{j}\right)-p_{c d}^{(i)}\left(a_{j}\right)\right|<\epsilon$  \textbf{Or} $i=i_{max}$}
  \STATE \textbf{Break} 
  \ELSE
  \STATE Go to step Procedure 
  \ENDIF


\item Outputs:

  \item $\hat{x}_{c}=\underset{a_{j} \in \mathbb{A}}{\arg \max } p_{c}\left(a_{j}\right), \quad c \in\{0, \cdots, N M-1\}$
   
 }
\label{MPalgorithm} 
 \caption{MP detection }
 \end{algorithmic}
\end{algorithm}



 \begin{figure}
  \centering
\includegraphics[scale=0.65]{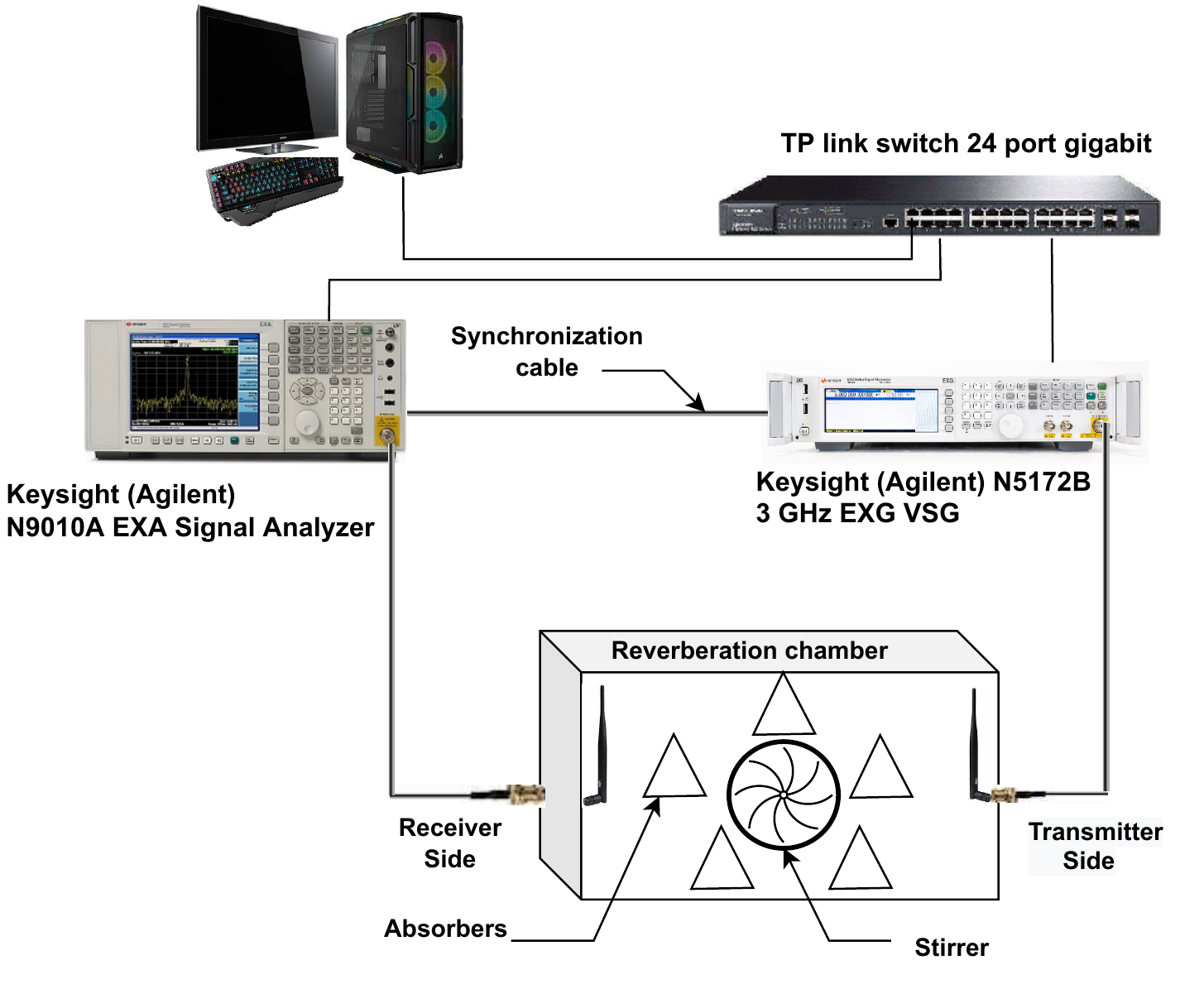}
  \caption{Laboratory equipment setup connection for multi-path emulation using Reverberation chamber.}
  \label{fig:Lab_equip_2}
\end{figure}
\section{Channel Effects And RF-Impairments}\label{Sec:Impairement}

\begin{figure}
  \centering
\includegraphics[scale=0.41]{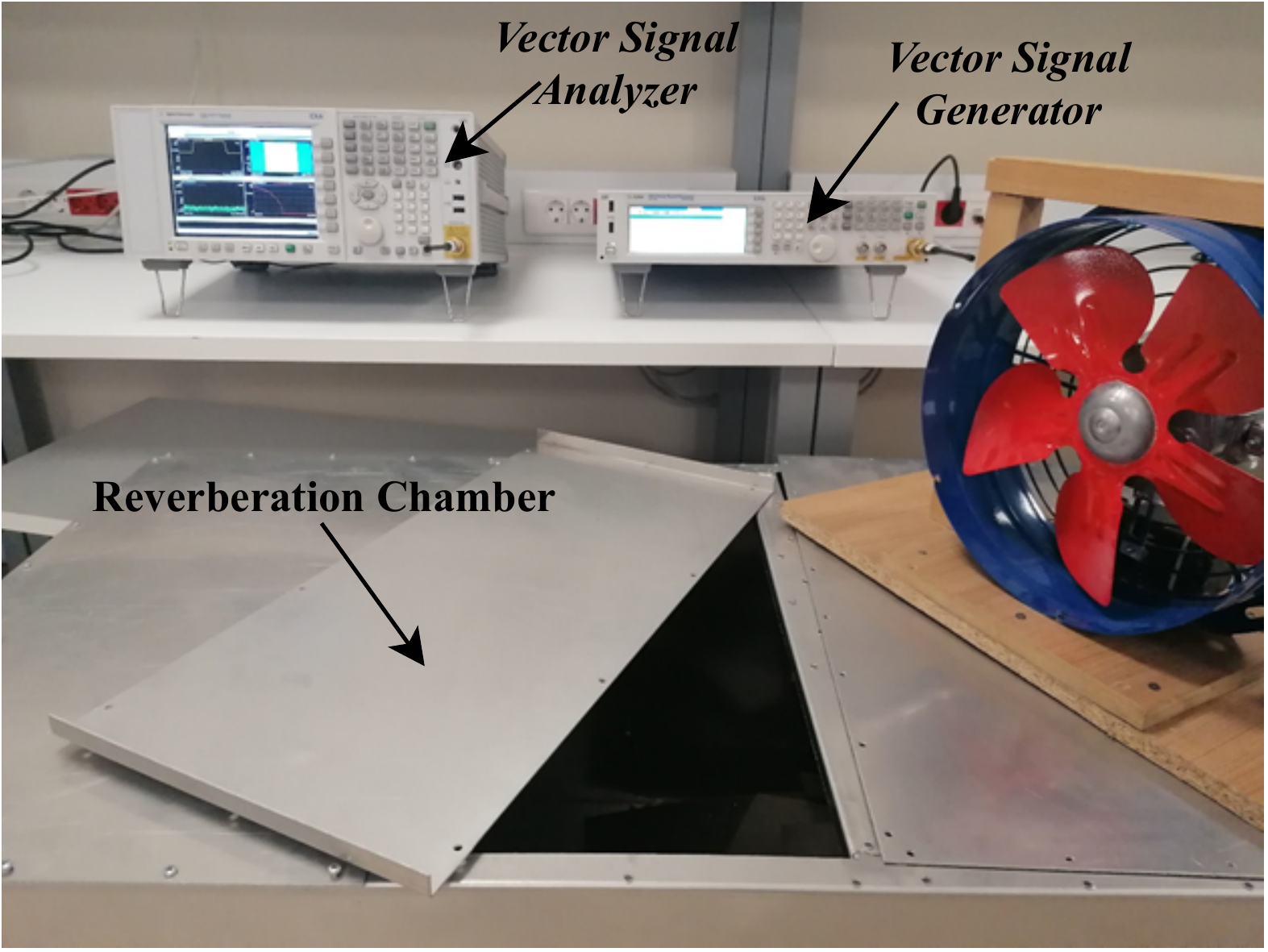}
  \caption{Laboratory equipment setup.}
  \label{fig:Lab_equip}
\end{figure}

The channel effect and the various RF-impairments degrade the system performance.Thus,in this section, the critical RF-impairments are discussed, and their effects on \ac{OFDM} and \ac{OTFS} are shown and compared experimentally. In the experiment platform, the SDR device based on the Keysight EXG X-Series design is used to implement terminals. As depicted in Fig. \ref{fig:Lab_equip}, the N5172B vector signal generator (VSG) with 9 kHz to 1, 3, or 6 GHz frequency range with output power +27 dBm, 900-µs switching speed. The receive antenna is connected to the N9010A EXA Keysight X-Series \ac{VSA} \cite{EXA}. Both the VSG and the signal analyzer are connected to the host PC running Matlab release 2020a software. The setup parameters are summarized in Table \ref{tab:Simulation Parameters}.

\begin{table}
\begin{center}
\caption{Simulation Parameters.} \label{tab:Simulation Parameters}
\resizebox{0.7\columnwidth}{!}{
\begin{tabular}{|c|c|c|c|}
 \hline
\textbf{Symbol}  &\textbf{Parameters}  & \textbf{Value (OTFS)}& \textbf{Value (OFDM)} \\
 \hline \hline 
 $f_c$  & Carrier frequency & $2.4GHz, 5GHz$ MHz & $2.4GHz , 5GHz$ \\
\hline
 $M$  & Number of subcarriers & $4,32,256$  & $256,~1024$  \\
\hline
   $N$  & Number of symbols & $4,32,256$ & $1$ \\
\hline
 $T_s$  & Symbol duration &\multicolumn{2}{c|}{$10~\mu s$}\\ \cline{3-4}
\hline
$M_{mod}$  & Modulation order & \multicolumn{2}{c|}{QAM}\\ \cline{3-4} 
\hline
 $\Delta f_s$  & Subcarrier spacing & \multicolumn{2}{c|}{$100$ KHz}\\ \cline{3-4} 
\hline
$f_o$  & Normalized frequency offset & \multicolumn{2}{c|}{$0,~0.05,~0.01,~0.1$}\\ \cline{3-4} 
\hline
 $\epsilon$  & I/Q gain imbalance & \multicolumn{2}{c|}{$0\%,~50\%$}\\ \cline{3-4}
\hline
 $\Delta \phi$  & I/Q phase imbalance & \multicolumn{2}{c|}{$0^o,~10^o$ degree}\\ \cline{3-4} 
\hline
\end{tabular}
}
\end{center}
\end{table} 

\subsection{ Carrier frequency offset }

In wireless communication, the propagating electromagnetic wave interacts with many obstacles called scatters, consequently, it is scattered, reflected, or refracted along with the propagation path. Therefore, for $L_{tap}$ different paths, $L_{tap}$ waves propagate with different delays and attenuation factors (i.e., multipath components). If the transmitter, receiver, or scatters are moving, these multipath components will be scaled in time equivalently causing a frequency shift (narrow-band) or frequency spreading (wide-band). 

\begin{figure*}[!t]
    \begin{center}
    \subfigure[Instantaneous spectrum
stirrer-off ]{\label{}\includegraphics[width=60mm,height=45mm]{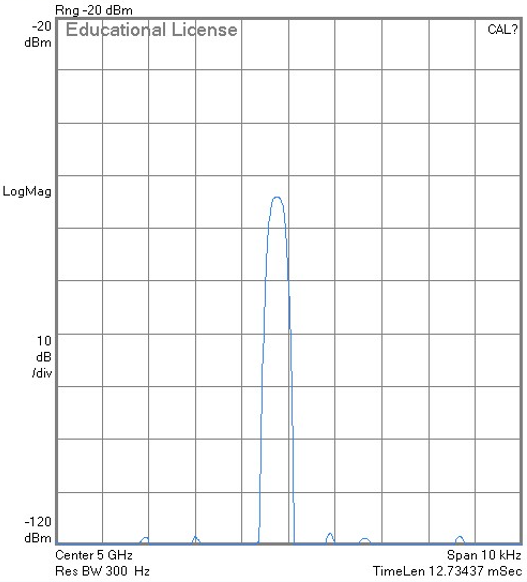}}
    \subfigure[Cumulative PSD when the stirrer-off]{\label{}\includegraphics[width=60mm,height=45mm]{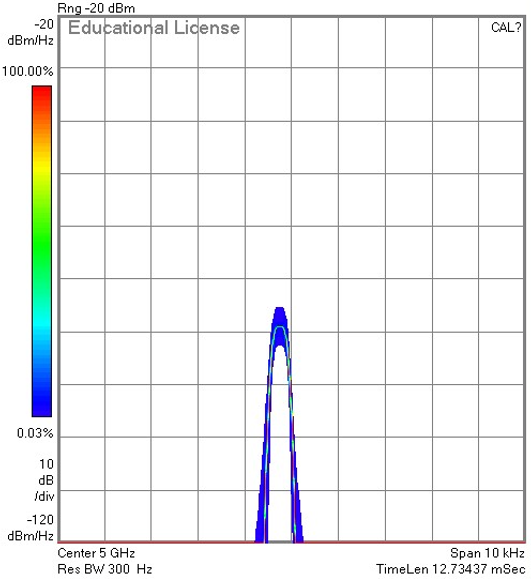}}
    \subfigure[Spectrogram when stirrer-off ]{\label{}\includegraphics[width=60mm,height=45mm]{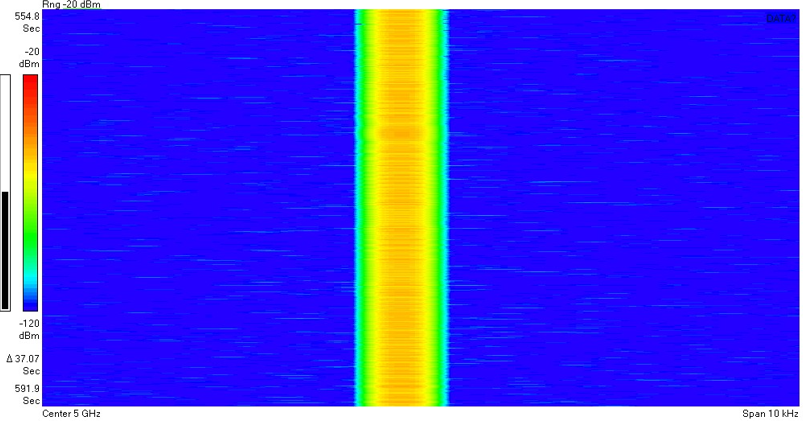}}
    \\
     \subfigure[Instantaneous spectrum
stirrer-on.]{\label{}\includegraphics[width=60mm,height=45mm]{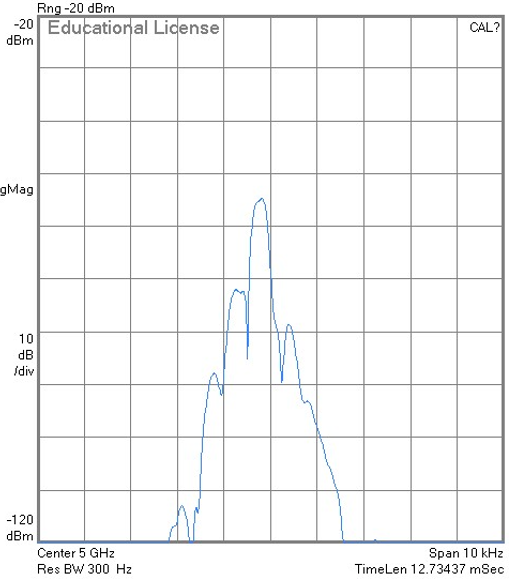}}
    \subfigure[Cumulative PSD when the stirrer-on]{\label{}\includegraphics[width=60mm,height=45mm]{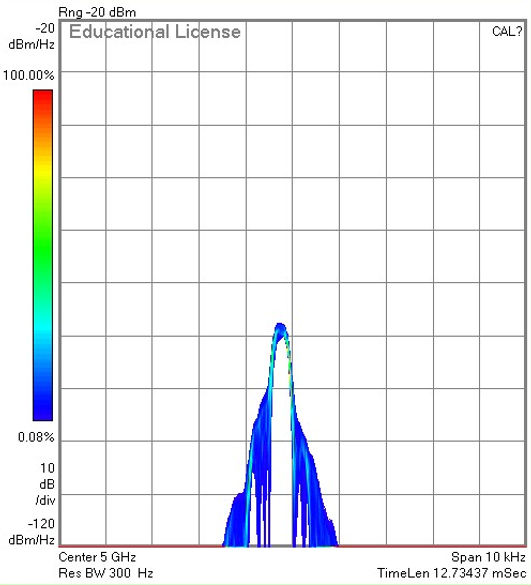}}
    \subfigure[Spectrogram when stirrer-on]{\label{}\includegraphics[width=60mm,height=45mm]{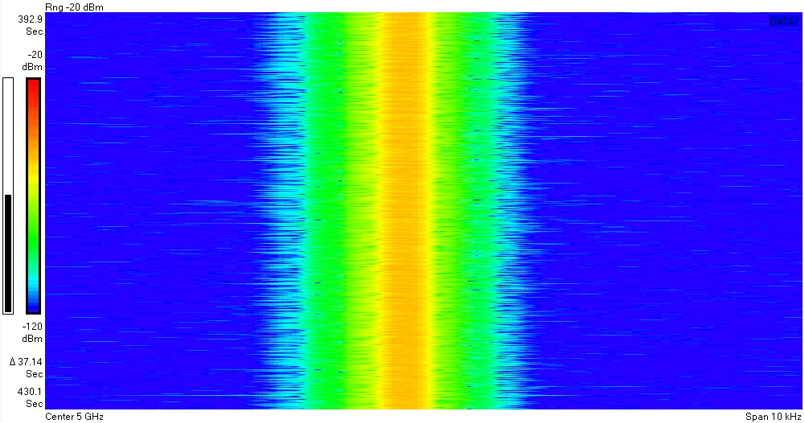}}
    
    \end{center}
    \centering
    \caption{Emulation the doppler shift effect on the 5GHz tone using reverberation chamber }
    \label{fig:Doppler}
\end{figure*}

\begin{figure} [h]
  \centering

\includegraphics[scale=0.5]{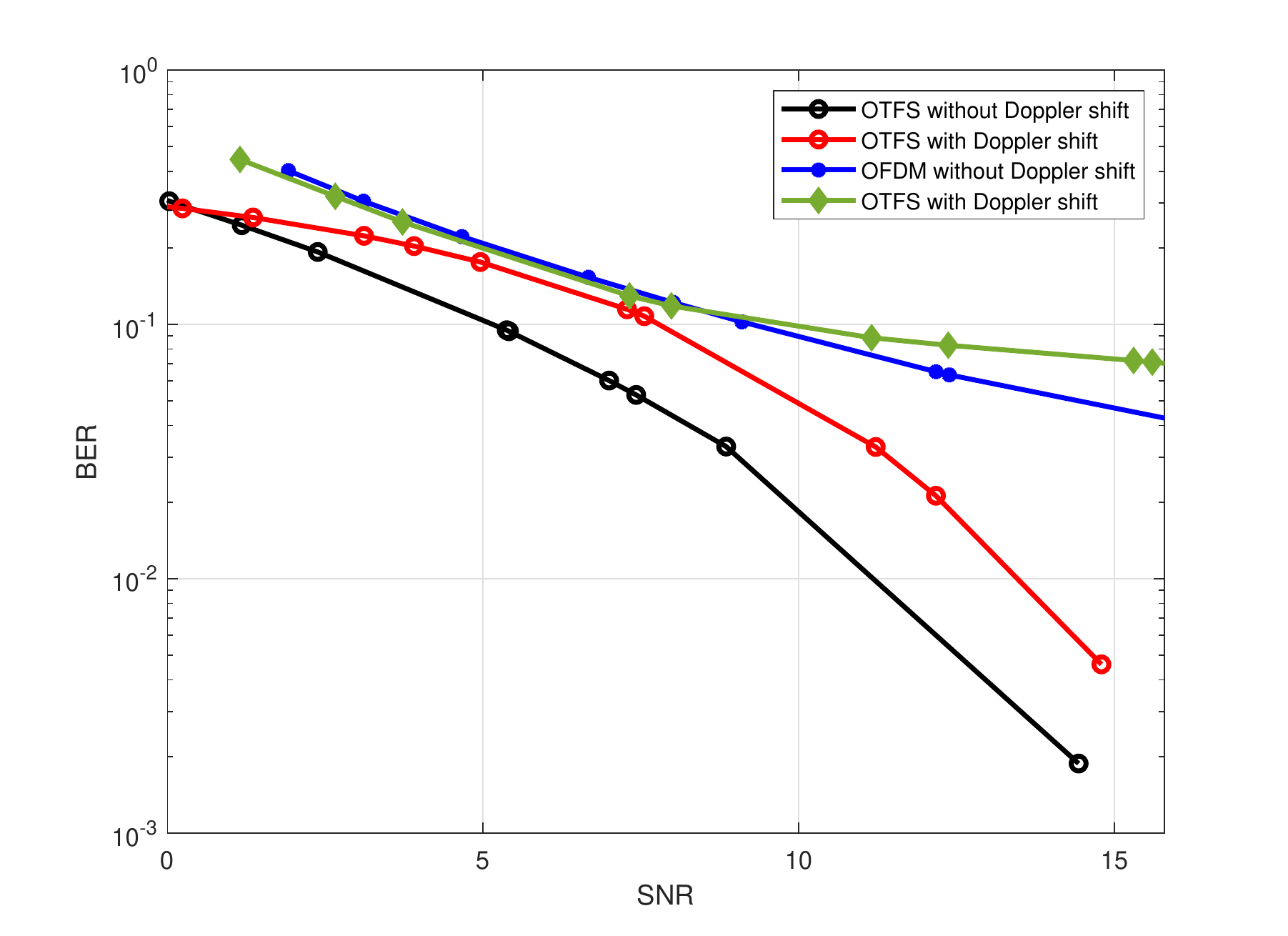}
  \caption{BER performance comparison between OTFS and OFDM effected by the Doppler shift using on the central frequency is 5GHz reverberation chamber.}
  \label{fig:cahmber}
\end{figure}

Thus, Carrier frequency offset (CFO) is occurred due to either Doppler effects or carrier frequency mismatching between the TX and RX oscillators\cite{mohammadian2021rf}.Therefore, we will discuss both CFO effect in our experiment.\par

Firstly, In our setup, to create Doppler effect a metallic fan (stirrer) with $40 cm \times 40 cm \times 20 cm$ dimensions with paddle radius of $0.15 m$; is used creating Doppler shifts for high speed. In order to convert these shifts into Doppler spread, a reverberation chamber with the dimensions of $ 120 cm \times 68 cm \times 55 cm$ is used to create more multi-path and provide enough Doppler spread for the experiment as shown in Fig. \ref{fig:Lab_equip_2} \cite{kihero2018emulation}. \par

Fig. \ref{fig:Doppler} illustrates the Doppler spread when a 5GHz tone is transmitted before and after turning the stirrer with the highest speed. Before turning on the stirrer, it is seen that the received signal has a constant bandwidth over transmission time, however, as soon as the stirrer is turned on the spectrum enlarges taking up more bandwidth due to the spreading.

When the Doppler shift is generated, the BER performance of both the OTFS and OFDM waveforms degrades, as seen in the Fig.\ref{fig:cahmber}. This degradation takes place as a result of the Doppler shift, which leads to the loss of orthogonality of the sub-carrier and eventually resulted in iter-carrier interference (ICI). However, as compared to OFDM waveform, OTFS waveform provides better BER performance. This is because the multipath components and Doppler shifts are resolvable in the delay-Doppler domain.
  
Assuming the resolved delays as $\tau_i$ and the Doppler frequency $\nu_i$, the received signal is given by the following weighted summation
\begin{equation}
y(t)=\sum_{i=1}^{L_{tap}} h_i x(t-\tau_i) e^{j 2 \pi \nu_i t} =\sum_{i=1}^{L_{tap}} h_i \left(\mathbb{M}_{\nu_i} \mathbb{D}_{\tau_i} x\right)(t),
\end{equation}
where $h_i$ is the $i$-th channel complex gain and $L_{tap}$ is the total number of resolvable paths.

Note that when using OTFS waveform, the multipath components and Doppler shifts are resolvable in the delay-Doppler domain, this is seen in Fig. \ref{fig:OTFS_signal} (c) where each bin represents a tap with specific delay and Doppler. This feature made the OTFS more suitable for rich scattering and high mobile environments than \ac{OFDM}.

Now, different frequency offsets have been implemented in the N5172B-VSG in order to replicate the carrier frequency mismatching that occurs between the TX oscillator and the RX oscillator. owing to the fact that the frequency offset of the N5172B-VSG might be adjusted.    

\begin{figure}
  \centering

  \includegraphics[scale=0.6]{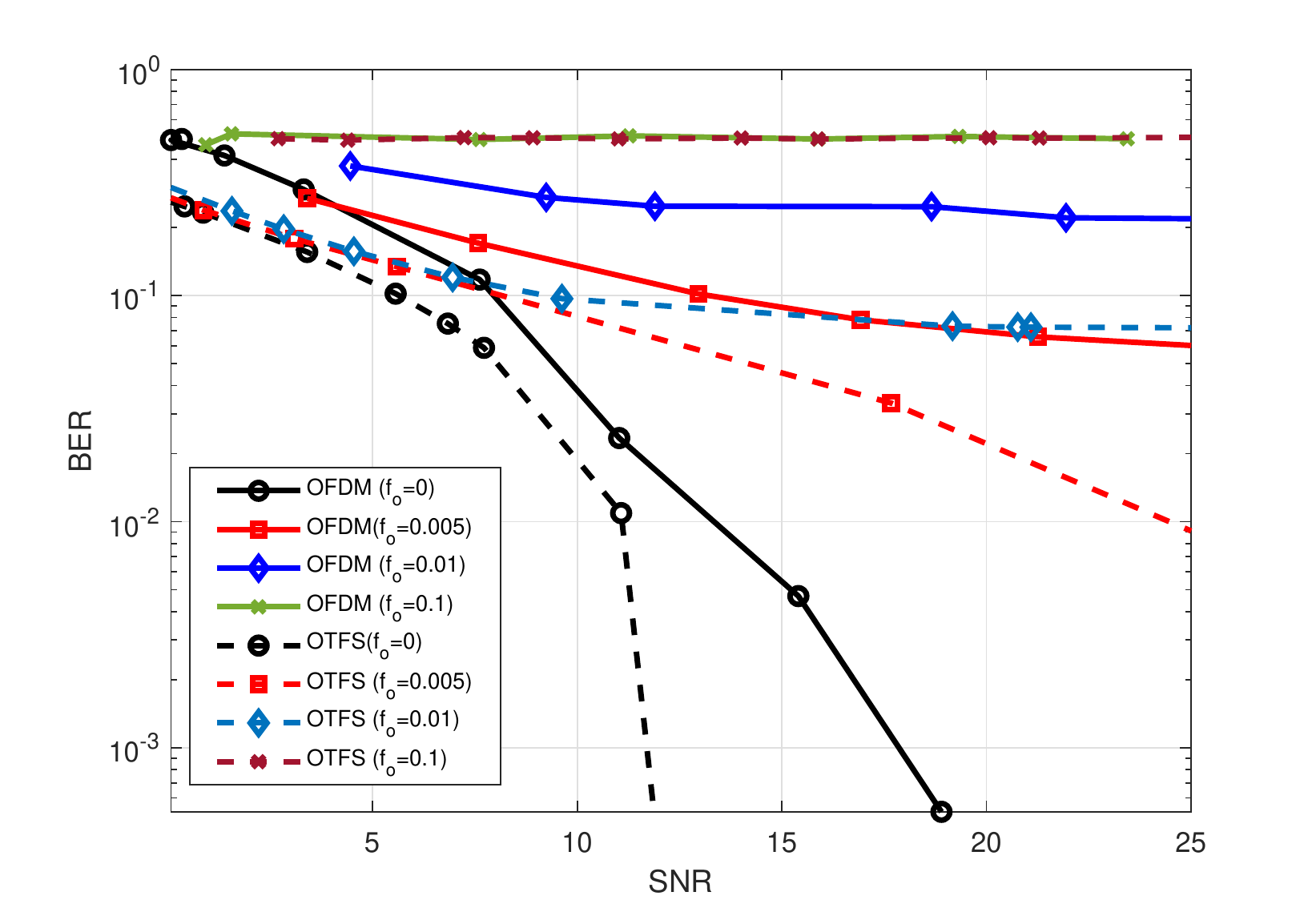}
  \caption{BER performance comparison between OTFS and OFDM systems with different normalized frequency offsets values in 2.4GHz carrier without using reverberation chamber.}
  \label{fig:CFO}
\end{figure}

The effect of the Doppler shifts caused by normalized frequency offsets may be seen in the BER performance of the OFDM and OTFS, which is presented in the Fig. \ref{fig:CFO}. As can be seen in the figure, the BER performance of both the OTFS and OFDM waveforms decays when there is an increase in the value of the mismatching frequency offset. In addition to this, it was demonstrated that OTFS has superior performance in comparison to OFDM on consideration of the reasons that have been discussed previously.
   
\subsection{Non-linearity Impairments}
Generally, there are different non-linearity sources in the RF front-end in communications systems namely, high power amplifier (HPA) at the transmitter, low-noise amplifier (LNA) at the receiver, mixtures, analog-to-digital (A/D) and digital-to-analog(D/A) converters[]. \par
In real-world wireless communication systems, HPAs are the most commonly used component to provide long-distance wireless transmission. According to the non-linear input-output characteristic of HPAs, the power of the input signal should be amplified within the HPA's linear range to prevent HPA to be saturated and cause the  out-of-band (OOB) that degrades the system performance[].   
The performance of the HPA amplifier is inversely proportional to the peak-to-average power ratio (PAPR) of the transmitted signal . As a result, the PAPR of the transmitted signal should be as minimal as possible \cite{hossain2020dft}.

The PAPR of the OTFS system is expressed as follow \cite{surabhi2019peak}.

\begin{equation}
{PAPR} =\frac{N \max _{k, l}|x[k, l]|^{2} }{\left.E\{\mid x[k, l]]^{2}\right\} }
\end{equation}
Due to the fact that the PAPR is a random variable, the best way to measure and  evaluate it is by using the complementary cumulative distribution function (CCDF) where the CCDF is presented as follows  \cite{surabhi2019peak}.

\begin{equation}
\begin{aligned}
\operatorname{Pr}\left(P A P R>\gamma_{o}\right) &=1-\operatorname{Pr}\left(P A P R \leq \gamma_{o}\right) \\
& \approx 1-\left(1-e^{-\gamma_{o}}\right)^{M N}
\end{aligned}
\end{equation}

where $Pr(.)$ denotes the probability function and $\gamma_o$ represents the threshold level that the PAPR of the transmitted OTFS signal should not exceed.

\begin{figure}[h]
  \centering
\includegraphics[scale=0.4]{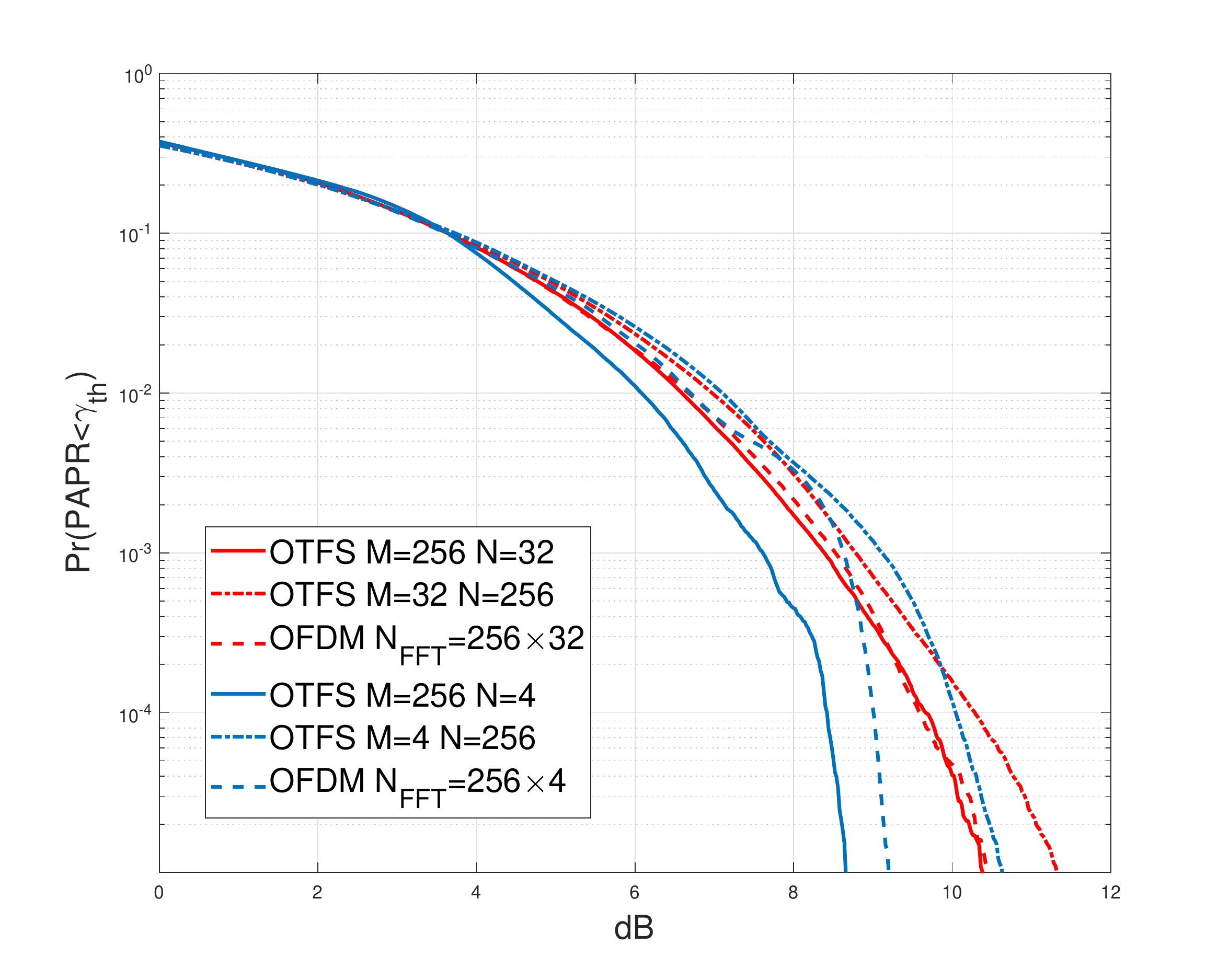}
  \caption{CCDF comparison of OTFS and OFDM.}
  \label{fig:PAPR}
\end{figure}
Fig. \ref{fig:PAPR} compares the CCDF of the PAPR for both \ac{OTFS} and \ac{OFDM} with the different values of $M$ and $N$ subcarriers. \footnote{Note that, for comparison of $M$-subcarrier OFDM with OTFS, which have $MN$ symbols in a frame, we consider the \ac{CCDF} of the concatenation of $N$ OFDM symbols.} It is shown that for the same number of sub-carrier where $M>N$, \ac{OTFS} provides better PAPR compared to \ac{OFDM}. For example, for $M=256$ and $N=4$, \ac{OTFS} has approximately 0.5 dB PAPR less than \ac{OFDM} at the probability of $10^{-5}$. Also, it is observed that as $N$ increases the PAPR of the \ac{OTFS} increases and the performance gap between the \ac{OTFS} and \ac{OFDM} almost vanishes. Note that in the case of $N>M$, it is shown that the PAPR of \ac{OTFS} is the worst among all, this was discussed in detail in \cite{PAPR}. 
 
\subsection{I/Q Imbalance}
In the communication system, direct-conversion receivers were used to convert radio frequency (RF) signals to base-band signals immediately. In contrast to heterodyne receivers, direct-conversion receivers did not need the signal to be down-converted to intermediate frequency (IF). Unfortunately, imperfections in the local oscillator (LO) might cause significant RF impairments, such I-Q imbalance and DC offset \cite{yih2009analysis}.\par
Local oscillator (LO) is a device that generates sine and cosine signals, which are used to represent in-phase and quadrature signals during the modulation and demodulation processes. To make a cosine signal, the LO produces a sine signal and shifts it $90^{\circ}$ degrees to produce a cosine signal. Unfortunately, the signal gain created and the phase $90^{\circ}$ are not in sync in terms of their practical implementation\cite{mohammadian2021rf}.
 \begin{figure}[h!]
  \centering
\includegraphics[scale=0.55]{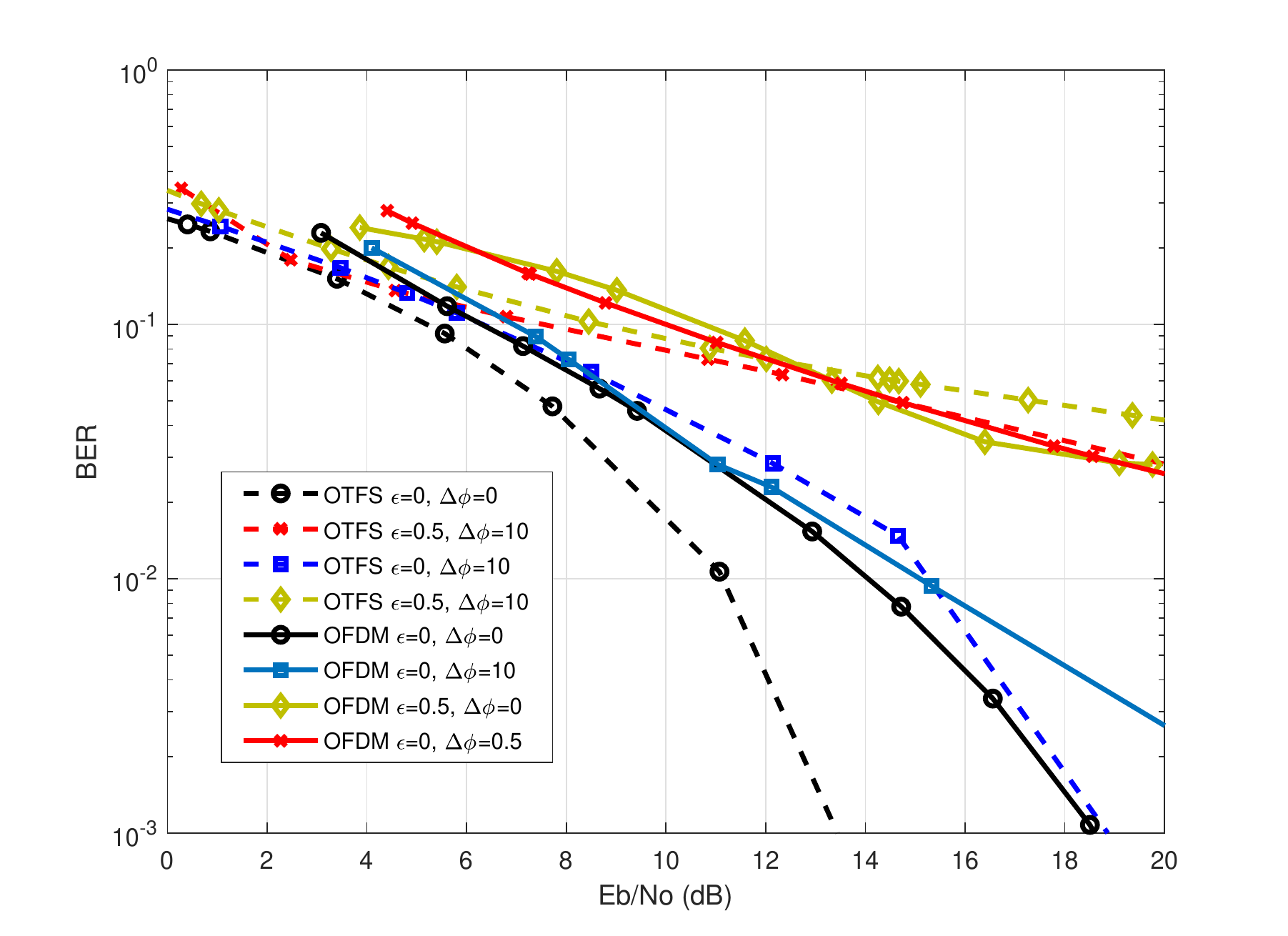}
  \caption{BER performance comparison between OTFS and OFDM systems with I/Q imbalances.}
  \label{fig:I/Q}
\end{figure}

\begin{figure*}[!t]
    \begin{center}
    \subfigure[$\epsilon=0,\Delta \phi=0$]{\label{}\includegraphics[width=60mm,height=40mm]{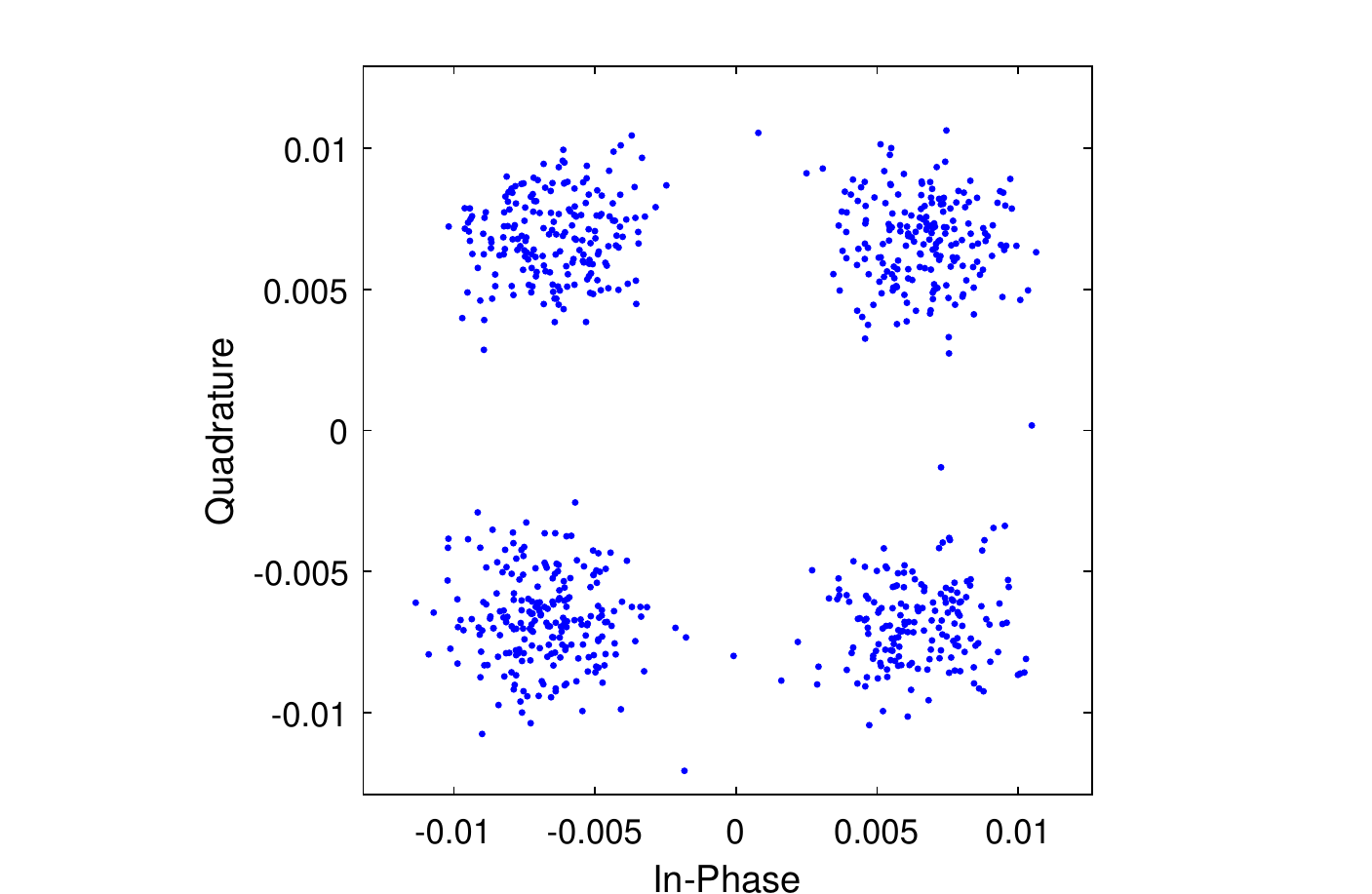}}
    \subfigure[$\epsilon=0,\Delta \phi=10$]{\label{}\includegraphics[width=60mm,height=40mm]{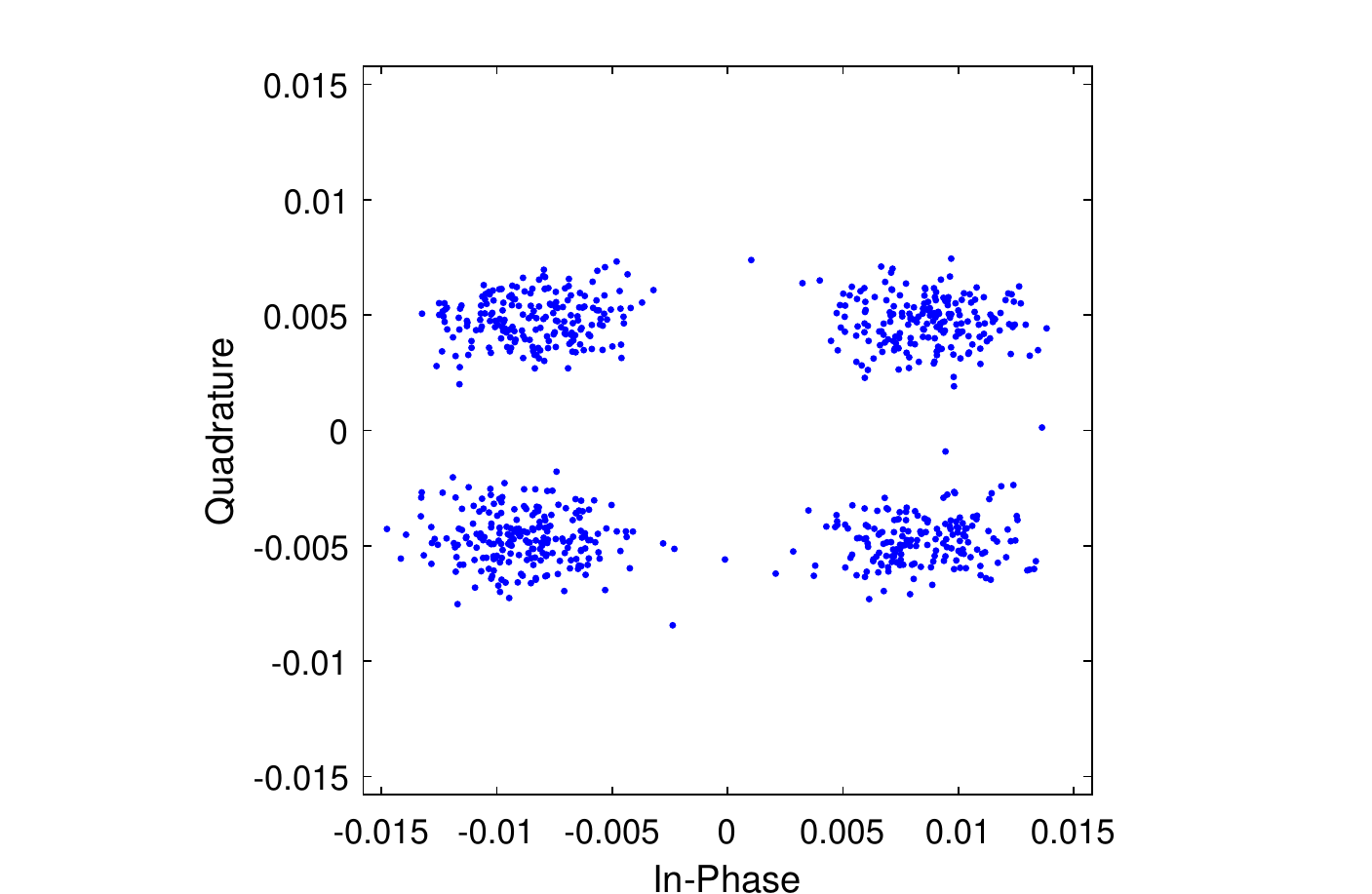}}
    \subfigure[$\epsilon=0.5,\Delta \phi=0$]{\label{}\includegraphics[width=60mm,height=40mm]{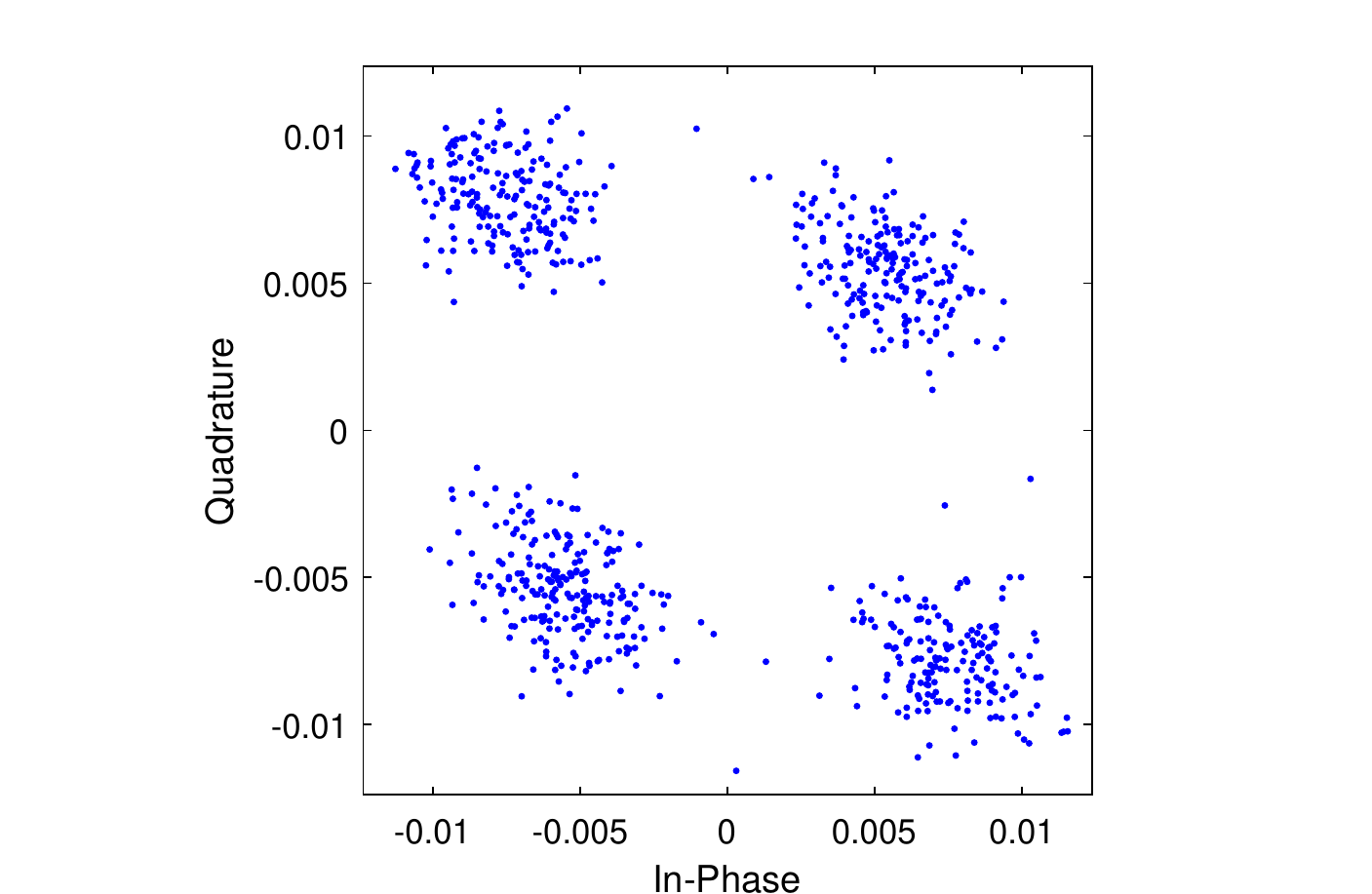}}
    \\
     \subfigure[$\epsilon=0.5,\Delta \phi=10$]{\label{}\includegraphics[width=60mm,height=40mm]{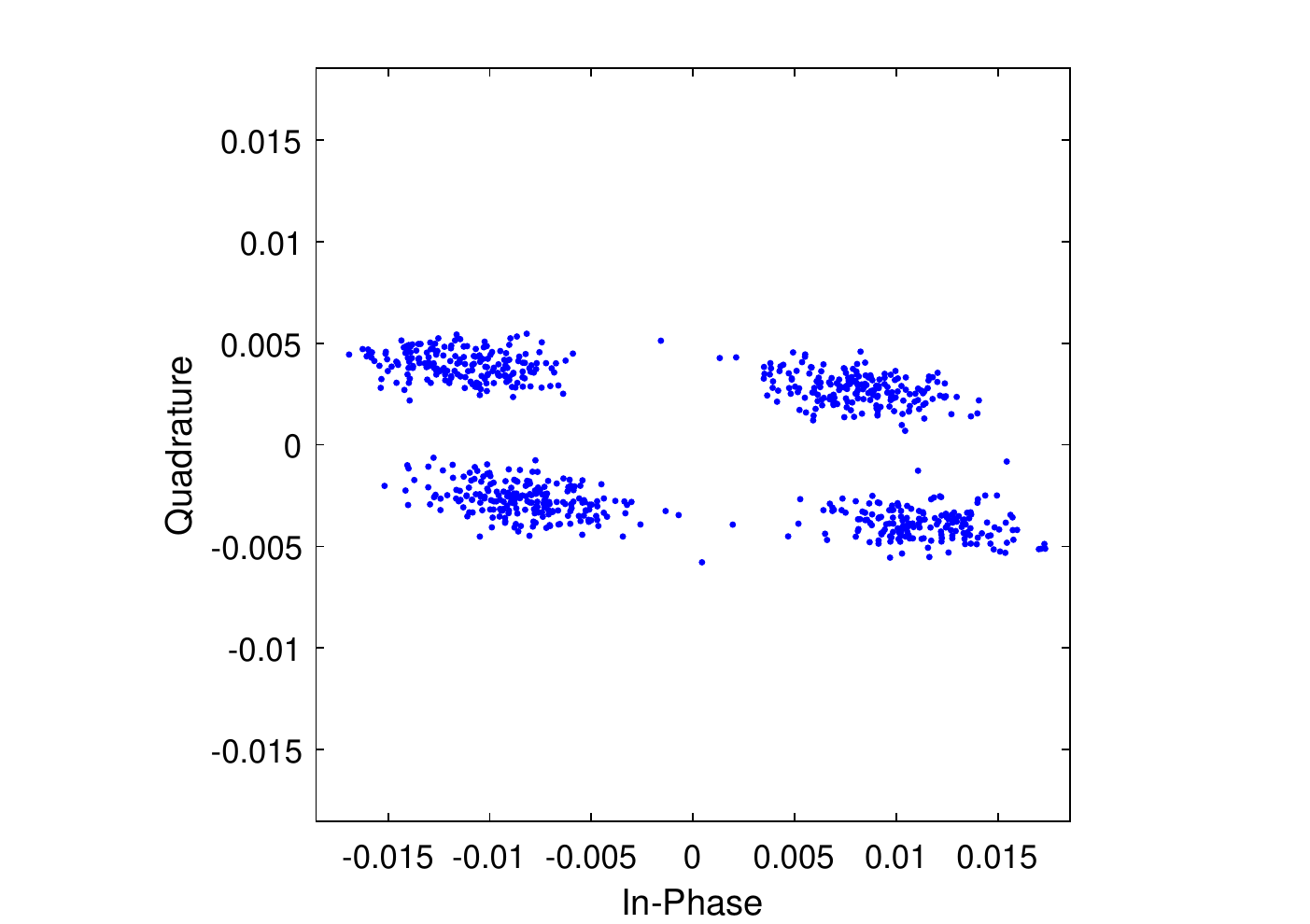}}
    \subfigure[$\gamma_I=0.75,\gamma_Q=0$]{\label{}\includegraphics[width=60mm,height=40mm]{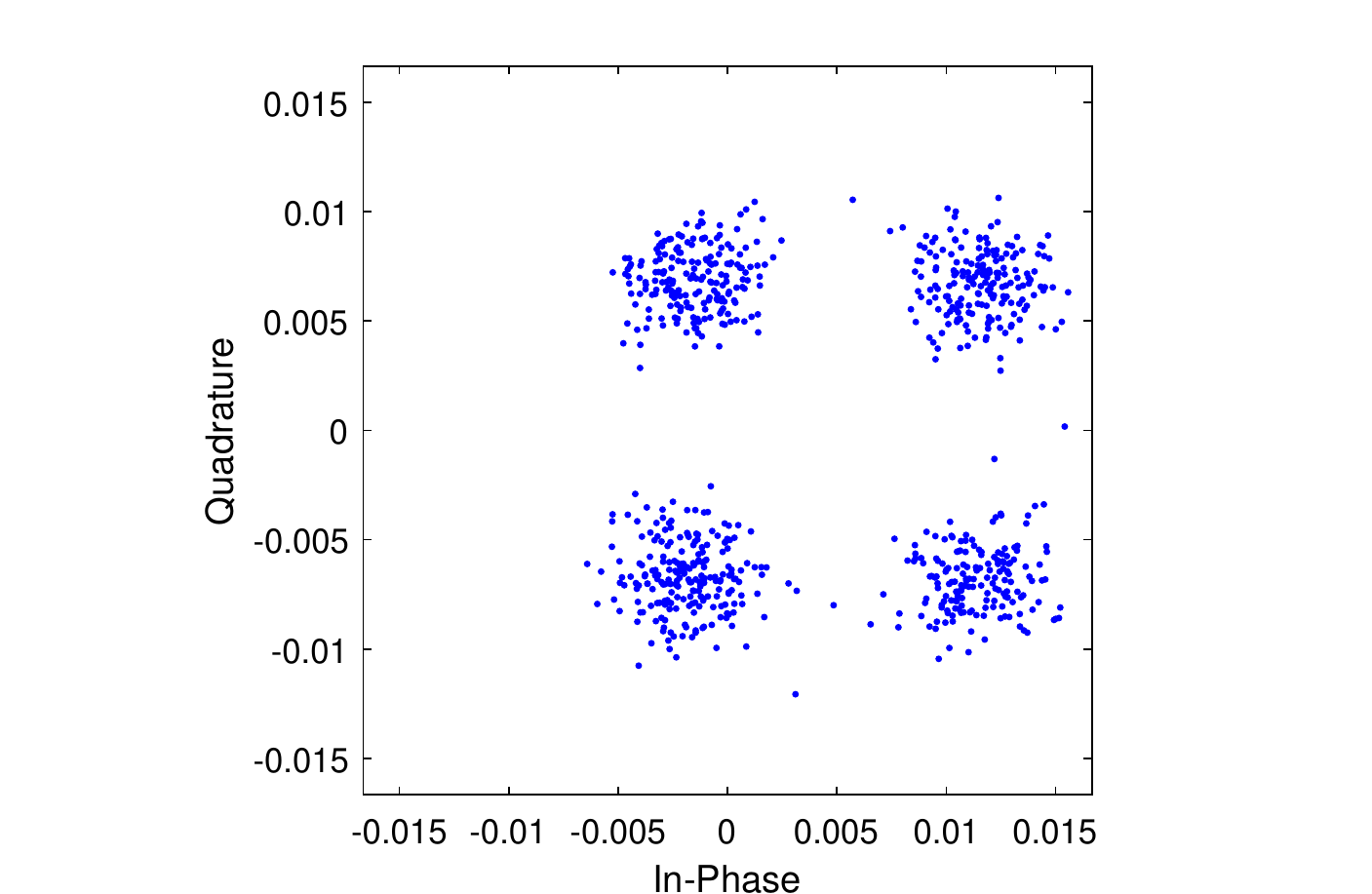}}
    \subfigure[$\gamma_I=0,\gamma_Q=0.75$]{\label{}\includegraphics[width=60mm,height=40mm]{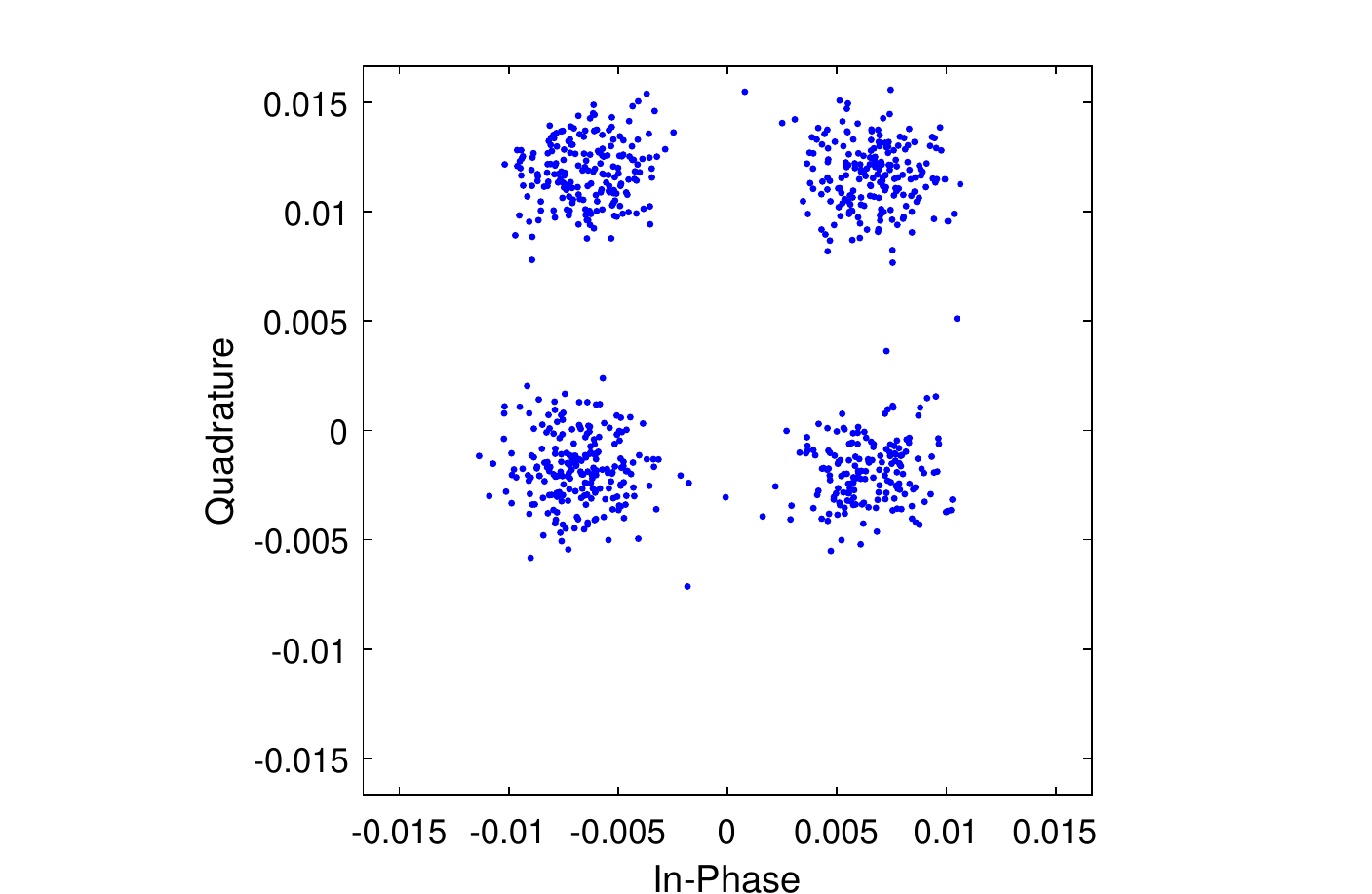}}
    
    \end{center}
    \centering
    \caption{Constellation diagram shows the effect of the different IQ and DC offset on the 4-QAM constellation diagram.}
    \label{fig:OTFS_signal}
\end{figure*}

I/Q imbalance impairment is the combination of two effects; the first one is the amplitude or gain imbalance (I) and the quadrature (Q) paths know as $\epsilon$, and the second one is the phase mismatch is given by $\Delta \phi$. 
The following model is used to add the I/Q imbalance on transmitted signal $x(t)=I+\jmath~ Q$ \cite{come2000impact,tubbax2003compensation}, as follows
\begin{equation}\label{I/Q_eq}
\begin{aligned}
y(t) &=(1+\epsilon) \cos \Delta \phi \Re\{x(t)\}-\jmath(1-\epsilon) \sin \Delta \phi \Re\{x(t)\} \\
&+\jmath(1-\epsilon) \cos \Delta \phi \Im\{x(t)\}-(1+\epsilon) \sin \Delta \phi \Im\{x(t)\},
\end{aligned}
\end{equation}where $\Re(\cdot)$ and $\Im(\cdot)$ symbolize the real and the imaginary part, respectively. For more simplicity, \eqref{I/Q_eq} could be written as 
\begin{equation} \label{I/Q_IM}
y(t) =\alpha \cdot x(t)+\beta \cdot x(t)^{*},
\end{equation}where $ (\cdot)^* $ denote complex conjugate, $\alpha=\cos \Delta \phi+\jmath \epsilon \sin \Delta \phi$, and $\beta=\epsilon \cos \Delta \phi-\jmath \sin \Delta \phi$. As it is observed in \eqref{I/Q_IM} that the I/Q imbalance does not exist if $\alpha=1$ and $\beta=0$.

In the N5172B-VSG , there is a specification for an internal I/Q baseband generator that may be adjusted either internally or externally, depending on the application. In this experiment, we adjust the internal I/Q baseband of the N5172B-VSG to evaluate the effect of I/Q imbalance on the performance of the OTFS system.

Fig. \ref{fig:I/Q} shows the effect of I/Q imbalance on the average \ac{BER} performance for both \ac{OTFS} and \ac{OFDM} systems. In general, I/Q imbalance degrades the system's performance for both waveforms as $\epsilon$ and/or $\Delta \phi $ increase, where as the I/Q imbalances are introduced in the system, the performance directly converges to a constant error floor at certain SNR value. Beyond this value, even increasing SNR does not help in improving the BER performance as given in \cite{IQ}. 

Additionally, it can be seen that changing the value of gain $\epsilon$ causes a bad influence that is greater than that of $\Delta \phi $. This is because changing $\epsilon$ leads to narrowing the received symbols in constellations, which in turn leads to narrowing the thresholds regions's that the demodulation should use to distinguish the received symbols. In other words, changing $\epsilon$ causes a more negative effect than changing $\Delta \phi $, Since the points in the constellation are now quite near to each other. While changing $\Delta \phi $ causes symbols to shift  without changing on the distance between neighboring points.

\subsection{DC offset}
The DC offset is also caused by the imperfection of the LO in direct-conversion receivers. where it is induced due to the leakage self-mixing of LO and transistor mismatch in the RF components\cite{neelam2022analysis}. 
DC offset will result in a shift on the symbols used in the constellation diagram (I/Q plane), and this shift might occur on the I-component, the Q-component, or both of them \cite{yih2009analysis}.  

\begin{figure}[h!]
  \centering
\includegraphics[scale=0.6]{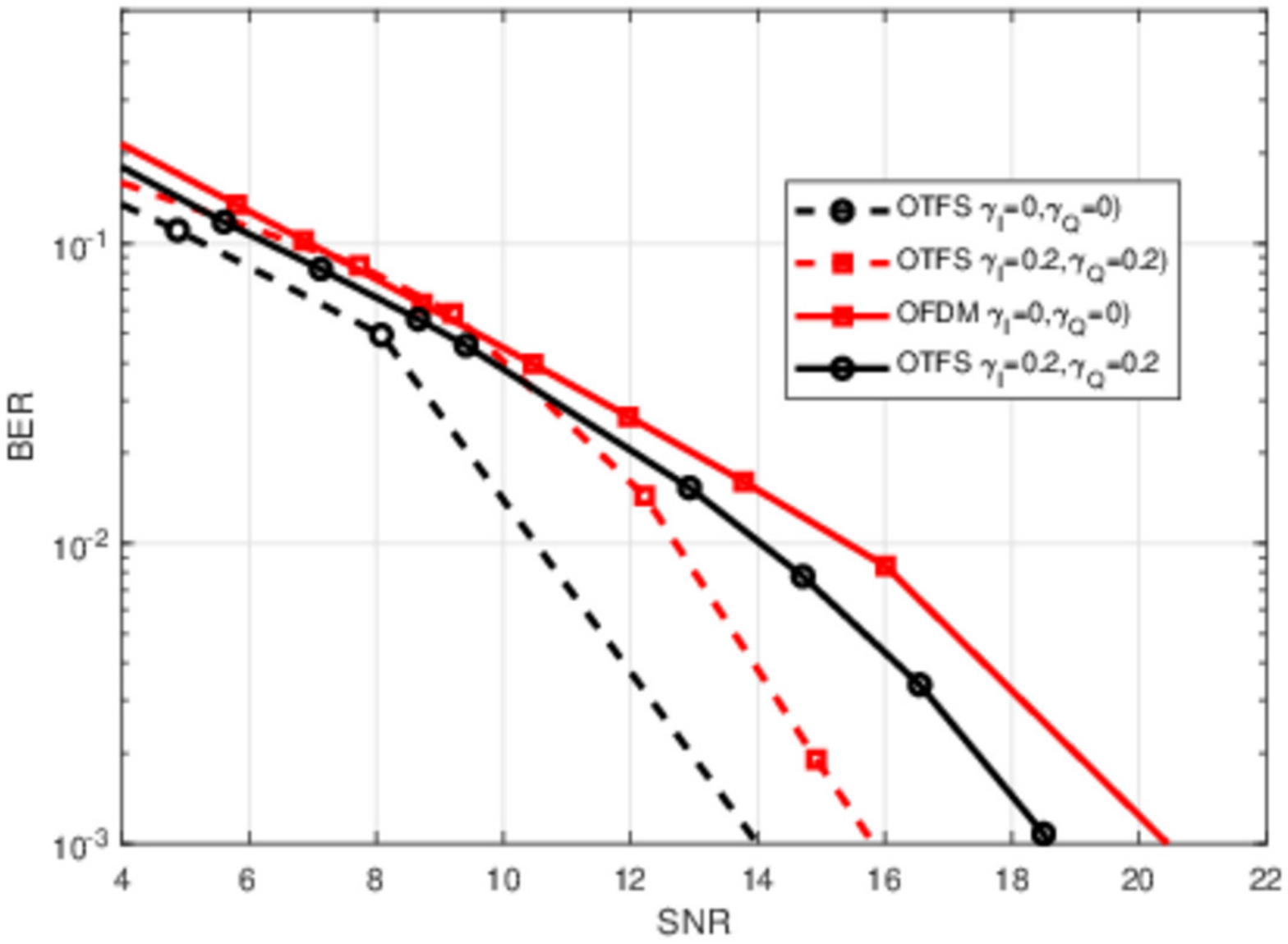}
  \caption{Comparison of the BER performance between OTFS and OFDM waveform}
  \label{fig:DC_OFFSET}
\end{figure}

The comparison of the impact of the DC-offset on the BER performance of OTFS and OFDM is shown in the Fig.\ref{fig:DC_OFFSET}, and it can be seen that the two waveforms have approximately the same influence on their BER performance. As was demonstrated, the effect of the DC-offset has a smaller impact on the system when the signal-to-noise ratio (SNR) is low, but it becomes more noticeable as the SNR rises. This is because the DC-offset expresses in the form of interference in the center of the transmission frequency spectrum, and as the SNR rises, the interference's impact on the system's overall performance accumulates.

\begin{figure}
  \centering
\includegraphics[scale=0.6]{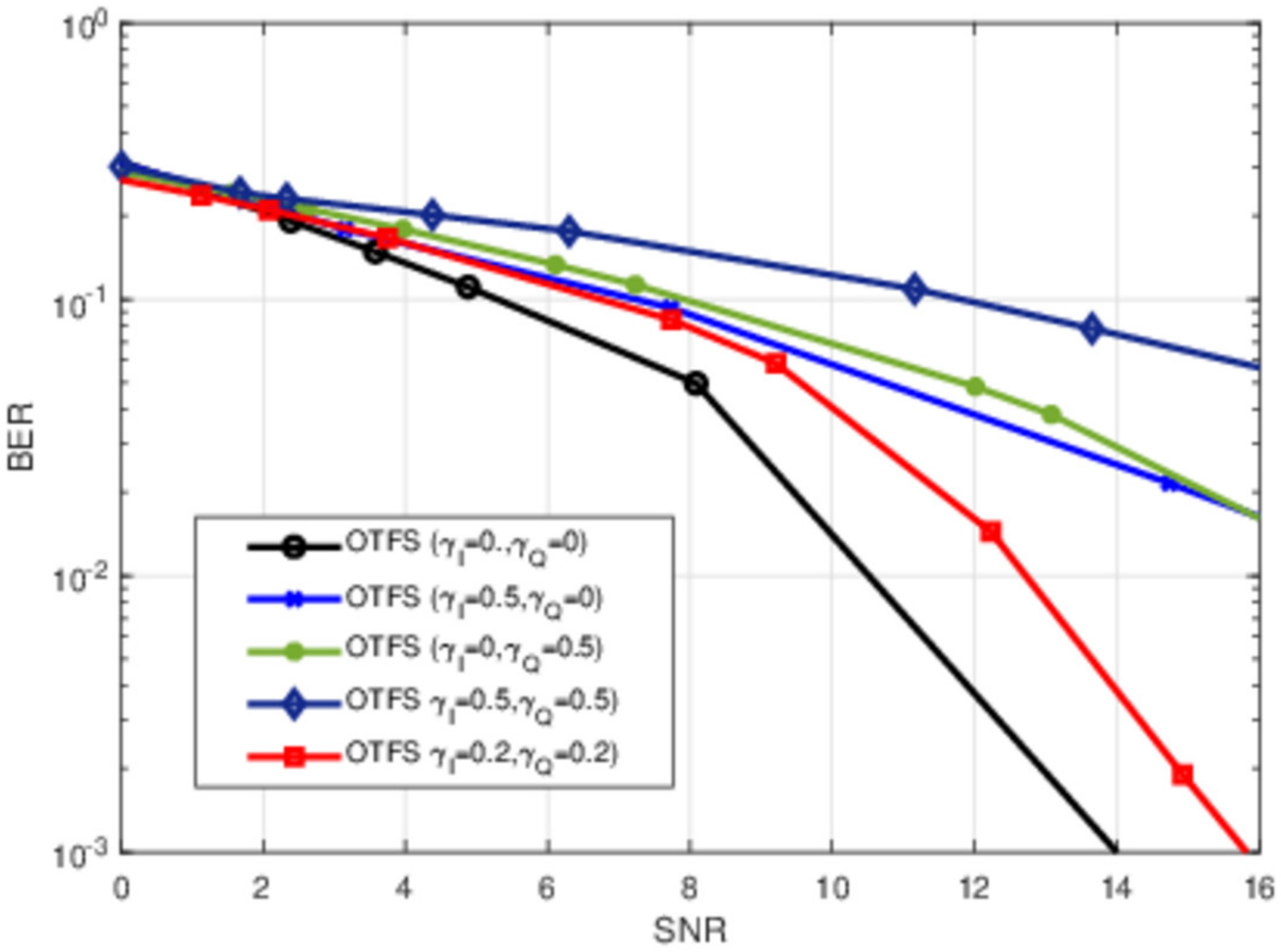}
  \caption{BER performance for OTFS under different values of DC-offset}
  \label{fig:DC_OFFSET_otfs}
\end{figure}

The impact that varying values of the DC-offset have on the performance of the OTFS waveform is seen in Fig.\ref{fig:DC_OFFSET_otfs}. According to the results, the degradation in system performance may be attributed to an increase in the DC-offset values. Additionally, it was shown that whether the in-phase DC-offset or the quadrature DC-offset, the system had the same influence on the performance of the system.

\subsection{Phase noise}

\begin{figure}
  \centering
\includegraphics[scale=0.7]{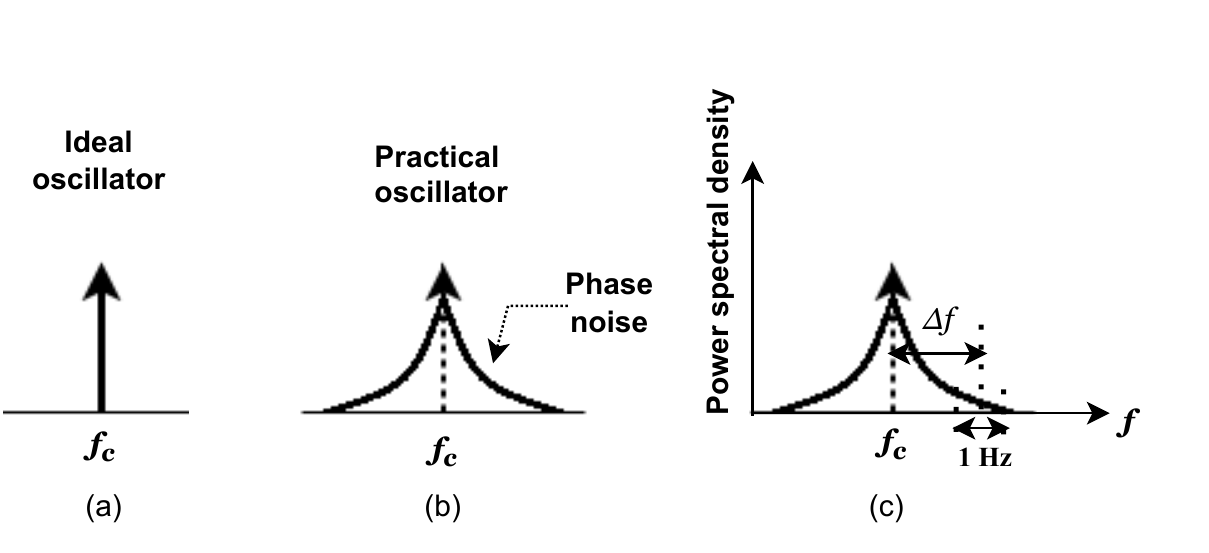}
  \caption{Phase noise of the local oscillator. (a) Ideal oscillator, (b) Practical oscillator, and (c) phase noise power level in dBc/Hz with $\Delta f$ offset }
  \label{fig:phasenois}
\end{figure}

When a local oscillator in a transceiver is unable to create pure sinusoidal waves in conformance with the Dirac spectrum, phase noise is produced as shown in Fig.\ref{fig:phasenois} (a) and (b) . The frequency spectrum and timing properties of the oscillator output induce large adjustments as a direct result of PN's influence \cite{mohammadian2021rf}.\par

In most cases, designers typically define PN in the frequency domain, using a bandwidth of one Hz and an offset of one $\Delta f$ from the carrier \cite{mohammadian2021rf}. The power of the PN signal throughout this bandwidth is normalized in relation to the power of the carrier in dBc/Hz unit as illustrated in Fig.\ref{fig:phasenois}(c).\par

\begin{figure}
  \centering
\includegraphics[scale=0.4]{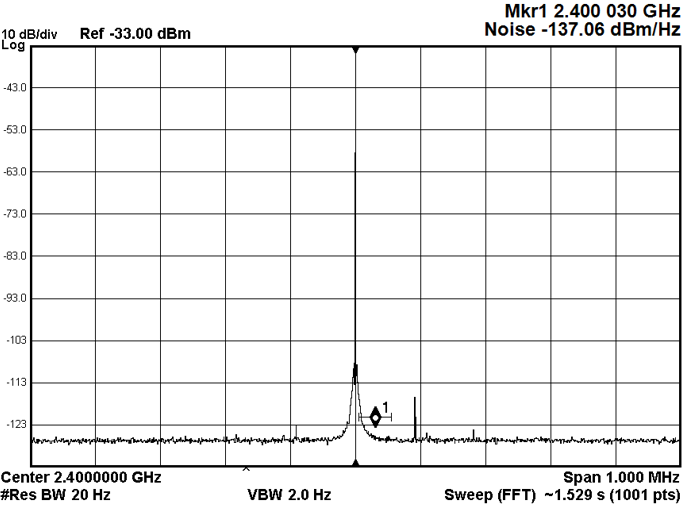}
  \caption{The phase noise of the VSA local oscillator}
  \label{fig:PN_30h}
\end{figure}

Both the N5172B vector signal generator (VSG) and the N9010A EXA Keysight X-Series vector signal analyzer (VSA) had amazingly low phase noise in our experiment.The phase noise of the VSA's local oscillator is shown in Fig.\ref{fig:PN_30h}, According to what been seen, the phase noise of the local oscillator is equivalent to -137.06 dBm/Hz at 30Hz frequency offset.\par

\begin{figure}
    \centering

\includegraphics[scale=0.5]{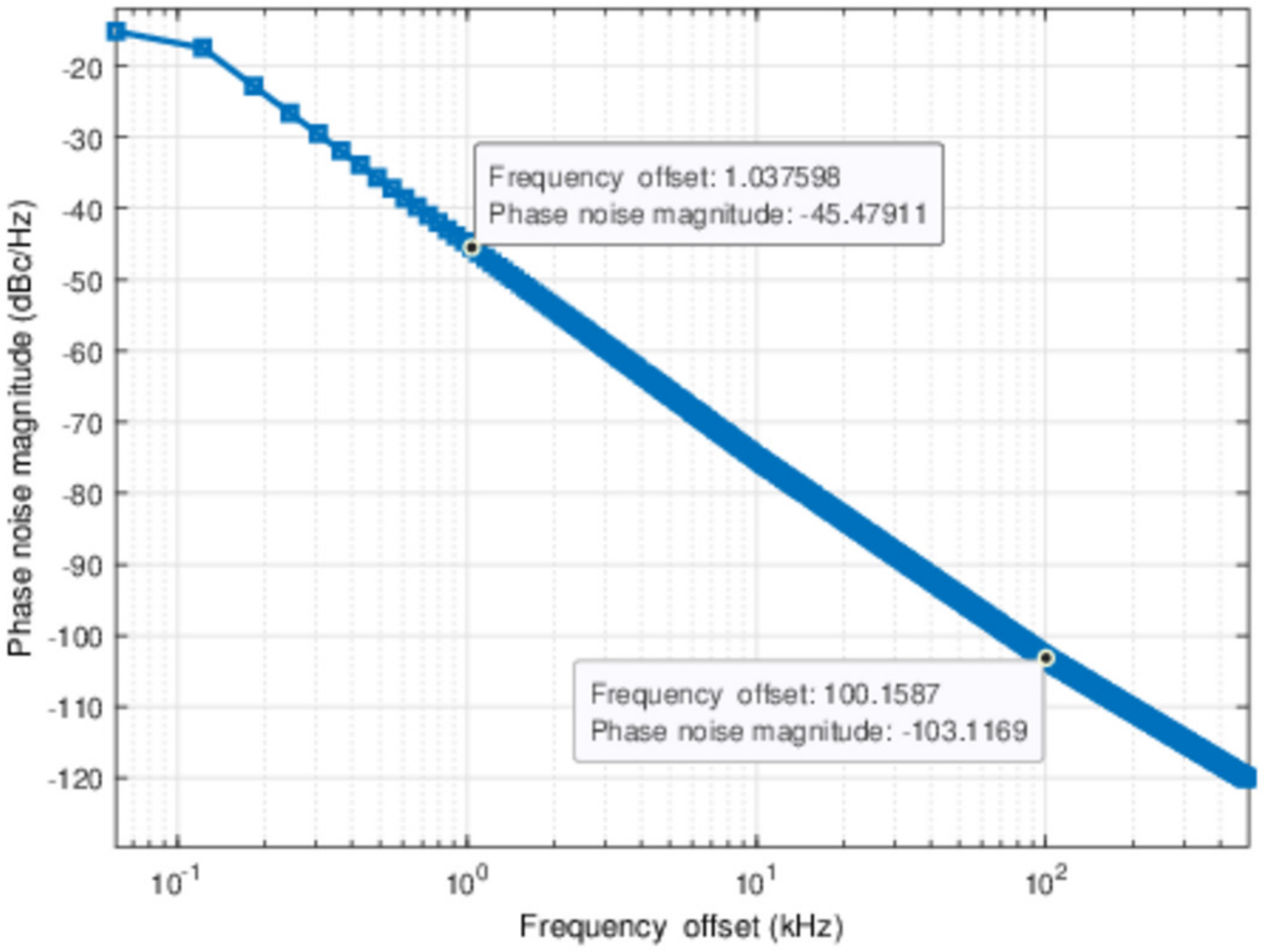}
  \caption{}
  \label{fig:PNmodle}
\end{figure}

The influence of phase noise will not be seen on the performance of the wave-forms systems because the level of phase noise is very low. Therefore, we have introduced phase noise at the receiver as illustrates Fig.\ref{fig:PNmodle}. And this modeling of the phase noise already exists on the 2.4GHz complementary metal-oxide-semiconductor voltage control oscillator (2.4GHz CMOS VCO) \cite{yan2008filtering}.

\begin{figure}
  \centering
\includegraphics[scale=0.6]{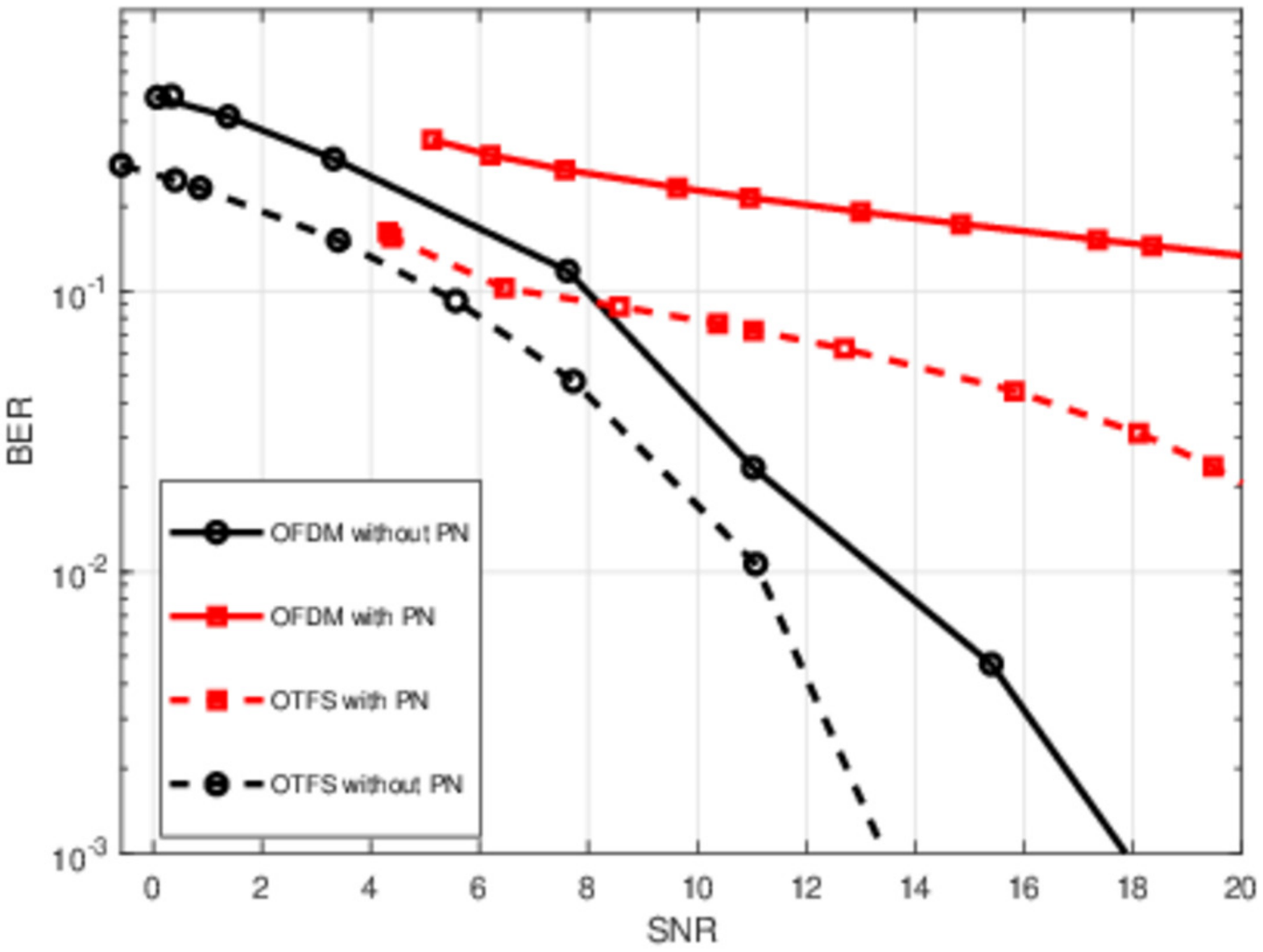}
  \caption{compersion of effect of the pahse noise on the BER performance for both OTFS and OFDM. }
  \label{fig:PN}
\end{figure}

The phase noise impact on OTFS and OFDM BER performance is shown in Fig.\ref{fig:PN} Phase noise has been proven to have a detrimental effect on both the system's performance and the orthogonality of the subcarriers, resulting in ICI. In contrast to OFDM, the OTFS waveform is more resistant to the phase noise effect.

 \section{Conclusion}\label{Sec:Conclusion}

This paper emulates and compares the RF-impairments of the OTFS waveform to OFDM impairments using SDR. The experiments were conducted in a real indoor wireless environment, where the metallic structure of the building introduced enough multipath, the Doppler shift was induced by the Stirrer, and the impairments were inherently produced inside the devices or added before transmission. The BER performance of the \ac{OTFS} modulation was superior than of the \ac{OFDM} under doubly dispersive channel. The CCDF of the PAPR of OTFS is shown to vary with the lattice structure in the delay-Doppler domain, and under $M>N$ condition, OTFS provides better PAPR compare to \ac{OFDM}. In addition, the I/Q-imbalance and DC-Offest impairments were explored, and the results showed that OTFS and OFDM are impacted in a manner that is approximately identical to one another. In addition, phase noise mitigation and OTFS give a higher level of phase noise resistance in comparison to OFDM. These findings provide an understanding of how and when to choose the waveform that is best appropriate for the characteristics of a particular channel.

\appendices

Appendixes, if needed, appear before the acknowledgment.
\bibliographystyle{IEEEtran}
\bibliography{bibliographie.bib}

\section*{Acknowledgment}

\end{document}